\documentclass[a4paper,11pt]{article}
\pdfoutput=1 

\usepackage{jheppub} 

\usepackage[T1]{fontenc} 

\DeclareMathOperator{\Tr}{Tr}
\DeclareMathOperator{\tth}{t\overline{t}h}
\DeclareMathOperator{\kt5}{\kappa_\xi^{\mathbf{5}}}
\usepackage{slashed} 			 
\graphicspath{{Figures/}} 	     

\title{\bf  Form Factors  in Higgs Couplings from Physics Beyond the Standard Model}

\author{Pedro Bittar} 
\author{and Gustavo  Burdman}
 \affiliation{Department of Mathematical Physics,\\
Institute of Physics, \\
University of Sao Paulo,\\
R. do Matao 1371, Sao Paulo, \\
SP 05508-090, Brazil}

\emailAdd{bittar@if.usp.br}
\emailAdd{gaburdman@usp.br}

\abstract{ We consider the momentum-dependent effects in Higgs couplings generated by physics beyond the standard model. We take a model-dependent approach, in which we can fully compute the non-local effects from physics not directly reachable by the LHC energy. We consider several scenarios, including composite Higgs models, additional scalars, and the continuum contributions of a quasi-conformal sector, as examples. For each specific model, we are able to obtain the form factor, with which it is then possible to fully simulate the effects in kinematics distributions. The momentum-dependent effects appear as a consequence of off-shellness in the process. We show how the sensitivity of different channels to the various models depends on how the flow of off-shellness appears in the Higgs couplings.

}

\begin{document} 
\maketitle
\flushbottom

\section{Introduction}
\label{sec:intro}
The discovery of the Higgs boson~\cite{Aad_2012,Chatrchyan_2012} completed the spectrum of the standard model (SM).
Furthermore, all available
experimental data seem to confirm so far the SM predictions for its
interactions, and bounds on masses of particles beyond the SM
spectrum have settled mostly above the TeV scale~\cite{Yuan:2020fyf}. Thus, despite its many
shortcomings, it appears that the
SM has been confirmed at this energy scale.  On the other hand, this
may not be completely surprising since most of the problems or
failures of the theory could be solved at much higher energies. Such
is the case, for instance, with the origin of flavor or neutrino
masses. Other problems, such as dark matter, for example, can be
addressed by sectors that are either very weakly coupled, very light or
both, resulting in not so stringent bounds from particle physics at
the TeV scale.  Traditionally, the hierarchy problem has been pointed
out as the one suggesting new physics at energies just above the weak
scale.  The tuning in the scalar mass squared growths quadratically
with the scale of new physics, the SM cutoff. With some notable
exceptions, most extensions of the
SM designed to solve the hierarchy problem are becoming   fine-tuned
due to the experimental bounds from searches at the LHC. This is the
case, for instance, for supersymmetric extensions as well as most
composite Higgs models since  both would imply the existence of
new particles the masses of which are significantly constrained by LHC
data. These bounds on new resonances sugest that future tests of these and other extensions of the SM
will come from precision measurements of the SM couplings at  higher luminosities, 
particularly once the high luminosity LHC (HL-LHC)
starts taking data.

It is in this context that we will consider deviations in the couplings of the Higgs boson. It is possible that these deviations
are not just constant shifts in the coupling  values but contain momentum dependence, leading to the presence of form factors.
These effects would result in deviations in kinematic distributions which, in addition to the deviations in total rates, would constitute important information on the new underlying dynamics. 
The appearance of momentum dependence in the Higgs couplings has been studied in a variety of contexts in the literature. 
For instance, a general treatment of these form factors was introduced in Ref. \cite{Bellazzini:2015cgj2}, where the form factors where also computed in a specific extension of the SM.  Ref.~\cite{Isidori:2013cla}, on the other hand, considered the associated production of a new scalar with the Higgs and a gauge boson, where the general treatment of the form factors was used to match to effective field theories (EFT) of BSM extensions, whereas in Ref.~\cite{Banerjee:2021efl} a general treatment of form factor effects in the presence of compositeness is presented. 

But the most common treatment of deviations in Higgs couplings, including momentum dependent ones, is through the use of EFT, where  all the possible higher dimensional operators consistent with the SM gauge symmetries are considered. There are two main approaches. In the SMEFT~\cite{Buchmuller:1985jz,Grzadkowski:2010es},  the Higgs boson is part of an electroweak doublet and the symmetry is linearly realized. In the HEFT~\cite{FERUGLIO_1993,Alonso_2013,Brivio_2014,Alonso_2016,Falkowski_2019}, on the other hand, the Higgs boson is considered to be a gauge singlet and the symmetry is non linearly realized. Although these approaches have the advantage of being model independent, the number of operators needed, even if we only consider up dimension six,  CP even, baryon and flavor conserving ones, is still quite large.
 Furthermore, the computation of some observables in the EFT would formally require dimension eight operators \cite{Azatov_2017},  which would make the entire program much more difficult to carry out due to the large number of additional operators needed.  Finally, there remain ambiguities in the matching of BSM scenarios to the EFT~\cite{Brivio:2017vri,Criado:2018sdb,Isidori:2013cla,Cohen_2021}. In particular, momentum dependent effects might be sensitive to these ambiguities as is stated, for instance, in Refs. ~\cite{Englert:2019zmt,Englert:2014aca}. 

 In this paper, we consider momentum dependent effects in Higgs couplings  in connection with BSM physics. Our aim is to map the most general aspects of Higgs form factors and how these will impact Higgs phenomenology. As opposed to the EFT cases, our approach will be {\em model dependent}.  We will study a few typical scenarios leading to these effects in the hope to put forward the most important features of signals that can be pursued experimentally. The key factor for capturing momentum dependent effects in the Higgs couplings will be the off-shellness, either of the Higgs or of particles coupled to it. We will see that, depending of the example studied, the requirement of off-shell particles will result in different possible signals.  
The form factor approach is well suited for capturing non-local effects as well as for comparison with experiments and is at least complementary to the EFT program.  The model dependence question can be addressed by considering a variety of BSM motivations for the shape and scale of the form factors, which can then be more easily  compared with kinematic distributions obtained at the LHC, as well as future facilities.  This is in fact, the program for the rest of the paper as well as for future work.

The plan for the paper is as follows: in Section~\ref{sec:form1} we will make some general remarks about form factors in Higgs couplings. In Section~\ref{sec:mchm} we  
 consider, as a first example,  Composite Higgs Models (CHM). Although these generally involve narrow resonances
that can be seen directly at colliders, it is interesting to obtain these form factors to show the different types of momentum dependence that appear. It is also an interesting way of showing that the on shell (and constant) modification of couplings that these models typically produce does not constitute a complete picture of the effects in CHM when off shell states are considered. In Section~\ref{sec:other} we consider a scalar mixing with the Higgs boson, as well as the possibility of the effects of a continuum above a certain infrared scale, possibly originated in a (quasi) conformal sector. In Section~\ref{sec:coll}  we comment on the phenomenology at the LHC of these various examples. Finally, we conclude in Section~\ref{sec:conc}.

\section{Higgs Form Factors} 
\label{sec:form1}

Momentum dependent couplings appear in a variety of contexts. Here we are interested in the effects that come from physics beyond the 
SM, and therefore we exclude form factors induced by perturbative effects within the theory.  Although in many cases these form factors are associated with compositeness, such as is the case in hadronic physics, we will not necessarily make this assumption  about the Higgs couplings.
We specifically focus on the Higgs boson since it is a less tested sector of the SM, and perhaps also the least understood. It is behind the origin of electroweak symmetry breaking and it generates the electroweak mass scale which is not radiatively stable in the SM. Thus, it may be a portal to physics at higher energy scales.

The study of the momentum dependence of the Higgs couplings can be complimentary to other searches for physics beyond the SM. Although typically form factors are associated with the existence of new channels, such as new particle resonances, it is of interest to pursue this avenue of phenomenology independently. For instance, the new resonances may be difficult to observe at present or even future colliders.
These effects may appear first in precision probes of the Higgs couplings before we have energy enough to directly observe new particles presumably responsible for the deviations. Alternatively, as we will discuss in Section~\ref{sec:other}, non-resonant effects could be responsible for momentum dependent deviations.  
Thus, exploring the capabilities of the LHC in Run 3, the  HL-LHC or other colliders to be sensitive to these effects should be an important component of the continuing study of the Higgs boson and its properties. In the rest of the  paper, we explore various possible scenarios so as to map several different phenomenological outcomes. Ultimately we want to identify the best observables for searching for this kind of physics. 

Extensions of the SM affecting the Higgs couplings will typically also produce a Higgs form factor. We will fully  illustrate this in the various examples in the next sections.
In general, we can see that the effects in the {\em normalization}  of the Higgs couplings will be of order
\begin{equation}
  \frac{v^2}{M^2}~,
  \label{dhonshell}
\end{equation} 
where $M$ is the new physics scale. These effects appear already in on-shell Higgs couplings. However, off-shell Higgs couplings will be generally affected by
corrections of order of
\begin{equation}
  \frac{q^2}{M^2}~,
  \label{dhoffshell}
\end{equation}
where $q^2$ is the typical off shell momentum squared. As we will show below, both effects are at least comparable even for modest off shellness.
Thus, once experiments become sensitive to off-shell  Higgs couplings thorough observables testing this kinematic regime, the inclusion of the momentum dependence in the couplings will be necessary. To start, let us consider a generic Higgs coupling to a SM particle $X$. In a general extension of the SM, the Higgs couplings are defined  at zero momentum, so we have the effective momentum dependent coupling
\begin{equation}
  c_{h,X}(q,p) = c_{h,X}^{\rm SM} \,\kappa_X\,F(q,p)~,
  \label{ffdef1}
\end{equation}
where $c_{h,X}^{\rm SM}$ is the SM coupling, $\kappa_X$ is a zero-momentum deviation in the BSM extension, $q$ is the Higgs momentum, $p$ is the SM $X$ particle momenta, and the form factor $F(q,p)$ is normalized so that
\begin{equation}
  F(0,m_X)=1~.
  \label{ffnorm1}
  \end{equation}
  In principle, when considering on-shell Higgs observables, we see that (\ref{ffdef1}) already implies that there will already be a deviation due to the form-factor. However, if we always consider on-shell Higgs observables, these effects can always be absorbed into a zero-momentum deviation $\kappa'_X=\kappa_X\,F(m_h^2,p)$ (i.e. the form factor evaluated with the Higgs on shell). Then, the only way to disentangle the form factor effects in (\ref{ffdef1}) is to include measurements with significant off-shellness, either in the Higgs or the other particles in the vertex. This, in turn, is the main driver in the phenomenology of these momentum dependent effects, since distributions would increasingly deviate from the SM prediction  in kinematic regions  associated with further  off-shell couplings.

 The same effects in (\ref{dhonshell}) and (\ref{dhoffshell}) can in principle be captured in a model-independent EFT approach. For instance, using the SMEFT up do dimension 6 operators, the matching of the new physics effects encoded in (\ref{ffdef1}) with the EFT would result in contributions to a number of operators. A model-independent fit~\cite{Ellis:2018gqa} of the coefficients of these operators would be less constraining than the predictions of a model, which typically fix most correlations among them. Thus, with a given data set it will always be advantageous to consider as many models as possible, affecting different experimental channels. This procedure could have more statistical significance, weather it is as bounds or as deviations from the SM, and it would be complementary to the completely model-independent EFT approach.  
Our aim here is to identify  groups of models that  result in similar momentum dependent effects and therefore demand  a specific set of signals  in order to be sensitive to them. 
As we will see below, the form factor $F(q,p)$ can result in enhancement or suppression, depending on the possible existence and position of resonances, or even in their absence. In the following sections we  explore some illustrative examples. 

\section{An Example: Composite Higgs Models}
\label{sec:mchm}
As a first example, we consider composite Higgs models (CHMs)~\cite{Agashe:2004rs,Panico:2015jxa,Bellazzini:2014yua} to demonstrate the main features of momentum dependence in the Higgs couplings, since they allow us to construct the Higgs form factors explicitly. Additionally, they are well motivated BSM candidates, providing a solution to the hierarchy problem and are consistent with the current experimental bounds from the LHC~\cite{ATLAS:2022ozf,CMS:2022yxp}. While interesting models on their own, most of this section can be viewed  as an application of the general ideas discussed in Section \ref{sec:form1}.

The effective low energy theory of CHMs with only the SM degrees of freedom can be partially completed into the UV to include heavy resonances. These  are generally expected in any specific realization in which a new strongly interacting sector emerges to generate the Higgs as a composite state. We are interested in a non-local formalism that captures the momentum effects of these resonances on the Higgs couplings. These non-local effects are naturally encoded in form factors.

In CHMs, a global symmetry is spontaneously broken at a scale $f\sim {\rm TeV}$, resulting in a spectrum of (pseudo-) Nambu-Goldstone bosons (pNGBs) that includes the Higgs boson.
The spectrum at this scale can be divided into elementary and composite sectors,  with interactions between the two:
\begin{equation}
	\mathcal{L}=\mathcal{L}_{ES}+\mathcal{L}_{CS}+\mathcal{L}_{int}.
\end{equation}
Both the Higgs and the heavy resonances emerge as part of the composite sector. The rest of the SM spectrum is in the elementary sector. The interactions between the two sectors
results in the explicit breaking of the global symmetry,  which in turn generates a potential for the Higgs with a nontrivial vacuum expectation value, and results in EWSB~\cite{Bellazzini:2014yua,Panico:2015jxa}.

The Higgs couplings to elementary states are generated by the interactions between the two sectors, $\mathcal{L}_{int}$.
There are various CHMs. Here we concentrate on the minimal realization, the so called MCHM$_\mathbf{5}$~\cite{Agashe:2004rs} . Below, we derive the Higgs couplings in the MCHM$_\mathbf{5}$ and we will see that these proceed trough resonances. 
The situation is analogous to what happens in hadronic physics, where the elementary photon mixes with the $\rho$ resonance, resulting in a dominant contribution to the pion electromagnetic form factor.
In CHMs, form factors are generally induced by TeV resonances that mediate the interactions of the Higgs with the elementary sector.


We will use the MCHM$_\mathbf{5}$ as the simplest CHM consistent with experimental observations. A new strongly interacting sector generates both the Higgs as a pNGB and a spectrum of heavy resonances, at the scale $f$. The strong sector is realized through the non-linear sigma model with $SO(5)/SO(4)$ coset, and a spectator $U(1)_X$ symmetry, present  to account for the correct fermion hypercharge assignments when including SM states\footnote{Such additional $U(1)_X$ symmetry will often be omitted since it does not play an important role in the form factor discussion. We refer to \cite{Agashe:2004rs,Panico:2015jxa} for a more complete overview of the MCHM$_\mathbf{5}$.}. The four degrees of freedom of the Higgs doublet, $h^{\hat{a}}$, are described in the non-linear realization by the Goldstone matrix
\begin{equation}
U[\pi]=\exp\left(\frac{i\sqrt{2}}{f} h^{\hat{a}} T^{\hat{a}}\right),
\end{equation}
in which $T^A=\{T^a, T^{\hat{a}}\}$ are the generators of $SO(5)$ divided into the unbroken components $a=a_L,a_R$ of $SO(4)\sim SU(2)_L\times SU(2)_R$ and the broken ones $\hat{a}=1,\dots,4$. The explicit generators and some relevant relations are defined in Appendix \ref{sec:MCHM5gen}.

Starting with the Higgs couplings to fermions, we assume that the elementary-composite interactions between the SM fermions and the Higgs occurs through the partial compositeness hypothesis~\cite{Agashe:2004rs}. CHMs generally use fermion partial compositeness to avoid reintroducing the hierarchy problem in the renormalization group evolution of the theory, as well as minimizing flavor violation.
Elementary fermions are linearly coupled to the composite sector via a $d=5/2$ composite operator $\mathcal{O}_F$. In the IR, the $\overline{q}_l \mathcal{O}_F^L$ interaction induces mixings between the elementary fermions, composite resonances and the Higgs.

\begin{figure}[tbp]
\centering
\includegraphics[width=0.7\textwidth]{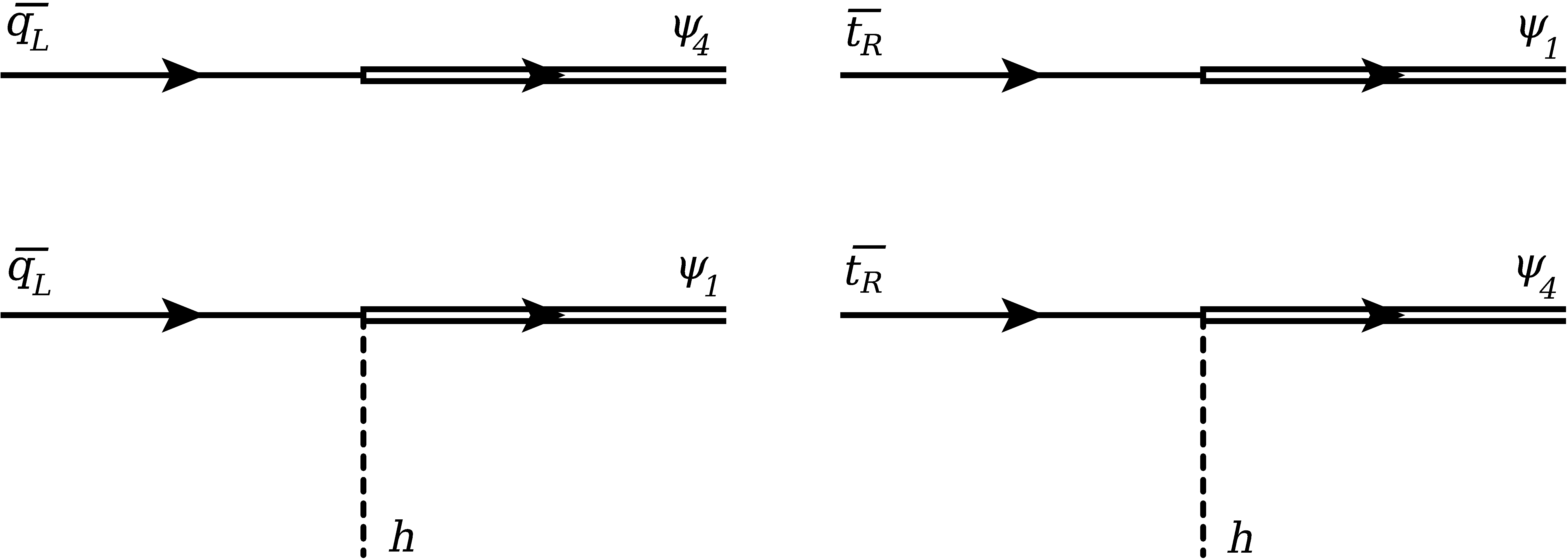}
\caption{Fermion vertices derived assuming the partially composite interactions in $\mathcal{L}_{int}$.}
\label{fig:ferm_vertices}
\end{figure}

In particular, in the MCHM$_\mathbf{5}$, the elementary  quarks are embedded as incomplete $\mathbf{5}$-plets of $SO(5)$. The left-handed fermions are identified as $\mathbf{4}$s of $SO(5)$ and the right-handed ones as $\mathbf{1}$s. For instance, for the third generation, we have
\begin{align}
q_L^{\mathbf{5}}=\begin{pmatrix}-i b_l\\ -b_l\\- i t_l\\ t_l\\0 \end{pmatrix} \qquad t_r^{\mathbf{5}}=\begin{pmatrix}
0\\ 0\\ 0\\0\\t_r \end{pmatrix} \qquad b_r^{\mathbf{5}}=\begin{pmatrix} 0\\ 0\\ 0\\0\\b_r \end{pmatrix}.
\end{align}  
  We will turn off the $b_{l,r}$ fields since the bottom quark will play no hole in our analysis. 
In order to proceed,  we  assume the existence of narrow $\mathbf{5}$-plet $\psi_i^{\mathbf{5}}$ resonances, in which the lightest ones dominate the interaction with quarks. Furthermore, the $\mathbf{5}$-plet resonances can be divided into fourplets and singlets of $SO(5)$. Then, the interaction lagrangian is given by
\begin{align}
\label{eq:ferm_int}
\mathcal{L}_{int}^{F}&\nonumber= f \left[y_{L1} (\overline{q_L^{\mathbf{5}}}U[\pi])_5 \psi_1 + y_{L4} (\overline{q_L^{\mathbf{5}}}U[\pi])_j \psi_{4,j} \right] +h.c.\\
 &\,\,+ f \left[y_{R1} (\overline{t_r^{\mathbf{5}}}U[\pi])_5 \psi_1 + y_{R4} (\overline{t_r^{\mathbf{5}}}U[\pi])_j \psi_{4,j} \right] +h.c.
\end{align}
where $j=1,...,4$, and we made use of the CCWZ procedure as described in Appendix \ref{sec:MCHM5gen}.
The induced Feynman rules can be obtained from \eqref{eq:ferm_int} by expanding the Higgs field, and are indicated in Figure~\ref{fig:ferm_vertices}.
\begin{align}
\label{ver1}& f y_{R1}(\overline{t_r^{\mathbf{5}}} U[\pi])_5 \psi_1 = f y_{R1} \cos\left(\frac{h}{f}\right)\overline{t_R}\psi_1 = f y_{R1}\overline{t_R}\psi_1+\mathcal{O}(h^2)\\
& f y_{L4}(\overline{q_L^{\mathbf{5}}} U[\pi])_j \psi_4 = f y_{L4} \cos\left(\frac{h}{f}\right)\overline{q_L}\psi_4 =f y_{L4}\overline{q_L}\psi_4 +\mathcal{O}(h^2)\\
& f y_{L1}(\overline{q_L^{\mathbf{5}}} U[\pi])_5 \psi_1 = f y_{L1} \sin\left(\frac{h}{f}\right)\overline{q_L}\psi_1 =  y_{L1} \overline{q_L}h\psi_1 +\mathcal{O}(h^3)\\
& f y_{R4}(\overline{t_r^{\mathbf{5}}} U[\pi])_j \psi_4 = f y_{R4} \sin\left(\frac{h}{f}\right)\overline{t_R}\label{ver4}\psi_4 = y_{R4} \overline{t_R}h\psi_4+\mathcal{O}(h^3).
\end{align}
The terms \eqref{ver1}-\eqref{ver4} will be responsible for generating the elementary-composite fermion mixing and interactions with the Higgs. The mixing of the left handed top quark with the $\mathbf{4}$ resonance will be combined with the Higgs interaction with $\psi_\mathbf{4}$ and the right handed quark to generate the top Yukawa coupling. The same applies to the $\mathbf{1}$ resonance.
So we conclude that in the MCHM$_5$ the Higgs coupling to elementary fermion states is dynamically generated via the exchange of a single $\mathbf{1}$ or $\mathbf{4}$ resonance.

Similarly to the fermionic case, the EW gauge bosons are embedded into $SO(4)$ representations and the elementary-composite interactions generate the gauge boson interactions with the Higgs. The EW group is a gauged subgroup of $SO(4)\sim SU(2)_L\times SU(2)_R \subset SU(2)_L \times U(1)_Y$. The physical gauge fields are identified as 
\begin{equation}
A_{\mu}=A_{\mu}^a T^a = g W^{a_L}_{\mu}T^{a_L} + g' B_{\mu} T^{3_R},
\end{equation}
where $A_{\mu}^a$ are the $SO(4)$ fields identified with the unbroken generators.
To include massive vector resonances that mix the elementary and composite states,  we assume that the vector resonances are associated with the spontaneous breaking of a hidden local symmetry (HLS)\cite{Bando:1984ej,Bando:1987br} . This corresponds  to the dual description of a $\left[\mathcal{G}/\mathcal{H}\right]_{\text{Global}}$ coset non-linear sigma model with a $\left[\mathcal{G}\right]_{\text{Global}}\times\left[\mathcal{H}\right]_{\text{Local}}$ model realized linearly. The vector  resonances become  dynamical massive gauge fields at the scale $f$, where their kinetic term is generated by strong dynamics. This is very similar to the original application of HLS to the description of vector meson dominance (VMD) in hadronic physics~\cite{Sakurai:1960ju}.

Assuming the dominance of the lightest vector  resonances, a very convenient way of introducing the gauge elementary-composite interactions is through kinetic mixing. The construction is closely related to the electromagnetic-hadronic interactions of VMD.  In fact, there are two representations of VMD, one based on kinetic mixing and the other based on mass  mixing. Here we make use of kinetic mixing. We
show the equivalence of both choices in Appendix~\ref{sec:basis}.
\begin{figure}[tbp]
\centering
\includegraphics[width=0.65\textwidth]{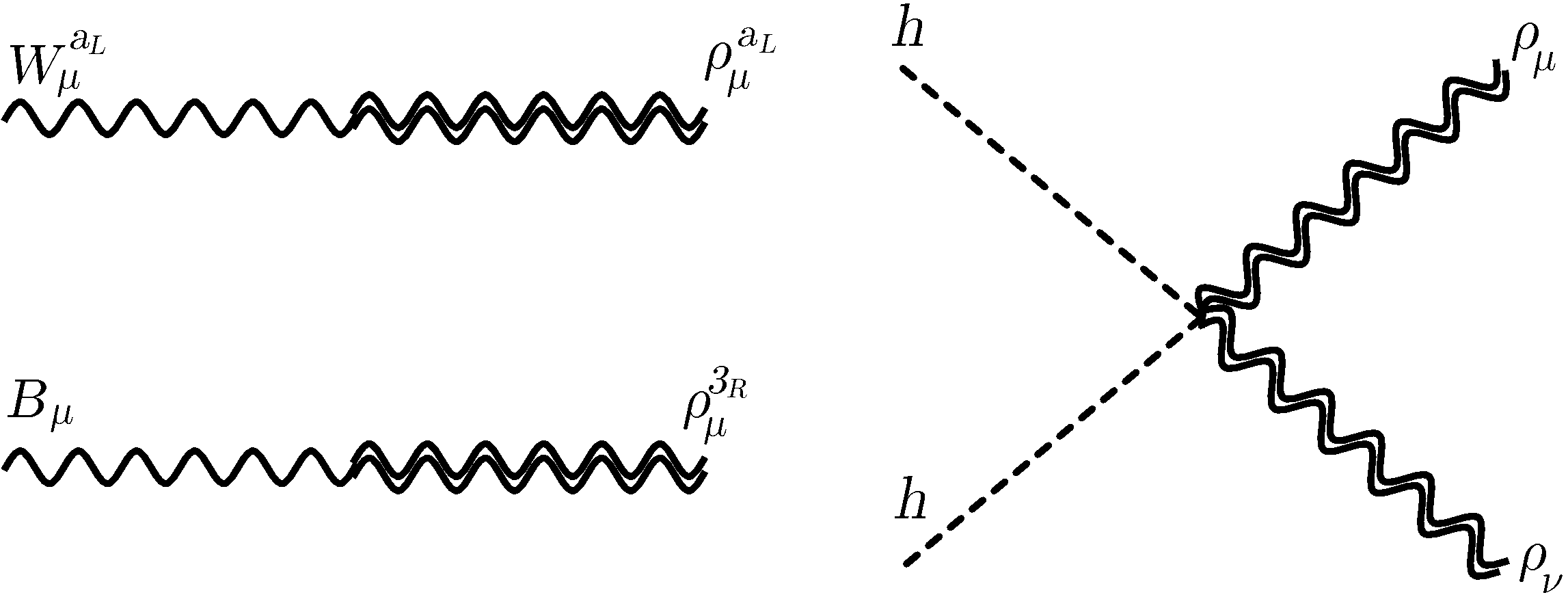}
\caption{Gauge bosons vertices derived assuming hidden local symmetry in the $\mathcal{L}_{int}$.}
\label{fig:gauge_vertices}
\end{figure}

\begin{figure}[tbp]
\centering
\includegraphics[width=0.65\textwidth]{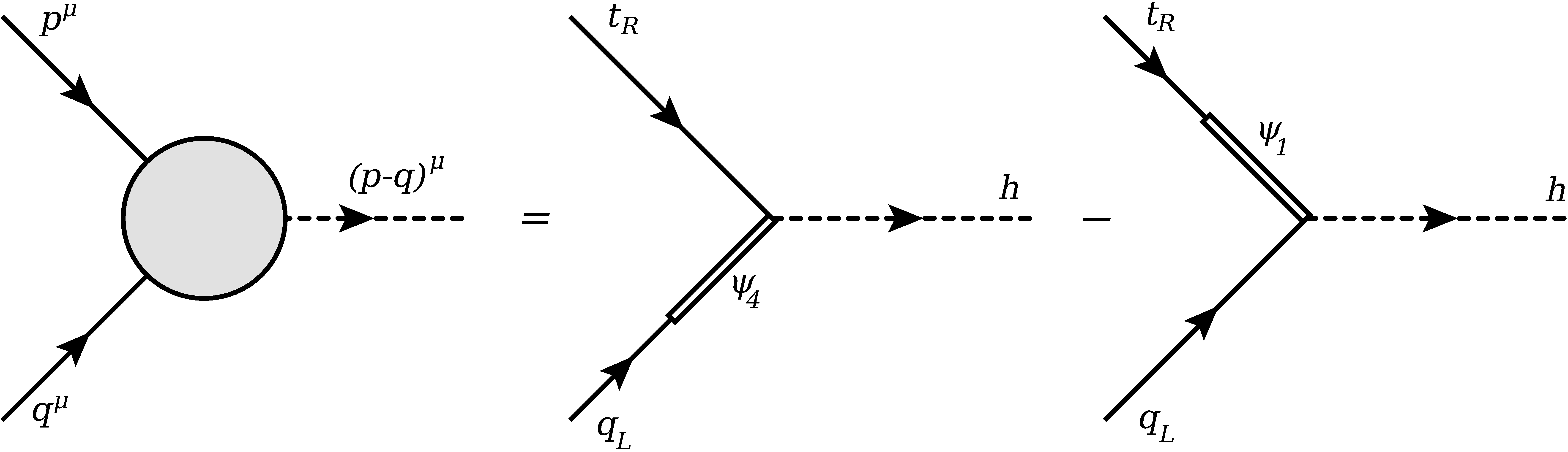}\\ \vskip0.5cm
\includegraphics[width=0.65\textwidth]{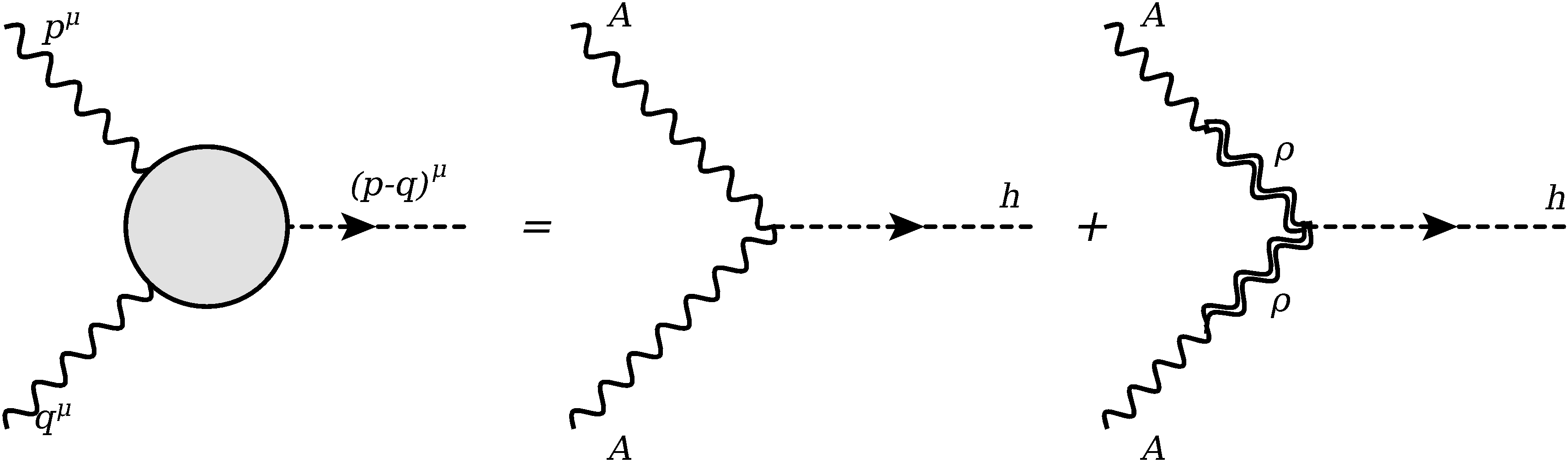}
\caption{\label{fig:higgs_couplings} Dynamical Higgs couplings to fermions and gauge bosons in the MCHM$_5$.}
\end{figure}

With the assumption of kinetic mixing, the lagrangian for the vector composite and elementary sectors, as well as for their interactions, is
\begin{align}
\mathcal{L}_{CS}^{V}=&-\frac{1}{4}\rho_{\mu\nu}^a \rho^{a,\mu\nu}+\frac{m_\rho^2}{2}\rho^a_{\mu}\rho^{a,\mu}+g_\rho \rho_\mu^a J^{a,\mu}+\frac{g_\rho^2}{2}\rho^a_{\mu}\rho^{a,\mu} h^2,\\
\mathcal{L}_{ES}^{V}=&-\frac{1}{4}W_{\mu\nu}^{a_L} W^{a_L,\mu\nu}+g W_\mu^{a_L} J^{a_L,\mu}+\frac{g^2}{2}W^{a_L}_{\mu}W^{a_L,\mu} h^2\\
&\nonumber -\frac{1}{4}B_{\mu\nu} B^{\mu\nu}+g' B_\mu J^{3_R,\mu}+\frac{g'^2}{2}B_{\mu}B^{\mu} h^2,\\
\label{eq:kinmix}\mathcal{L}_{int}^{V}=&\frac{1}{2}\frac{g}{g_\rho} W_{\mu\nu}^{a_L} \rho^{a_L,\mu\nu} + \frac{1}{2}\frac{g}{g_\rho} B_{\mu\nu} \rho^{3_R, \mu\nu},
\end{align}
Expanding the interaction term \eqref{eq:kinmix} in momentum space will lead to a momentum dependent mixing after integrating by parts.
\begin{equation}
\label{eq:kinmix2}\mathcal{L}_{int}^{V}=\frac{g}{g_\rho} q^2 W_{\mu}^{a_L} \rho^{a_L,\mu} + \frac{g'}{g_\rho} q^2 B_{\mu}\rho^{3_R, \mu}
\end{equation}
The Feynman rules obtained from kinetic mixing are shown in Figure~\ref{fig:gauge_vertices}. Gauge invariance forbids the appearance of Higgs interactions with a single vector  resonance. Instead, unlike in the case for fermions, the Higgs coupling to elementary vector states is dynamically generated by the exchange of two vector  resonances as indicated in Figure~\ref{fig:higgs_couplings}.

\subsection{Composite Higgs Form Factors}

In order to obtain momentum dependent effects in the Higgs couplings,  we  integrate out the resonances. The resulting  non local effective theory  maintains the correct pole structure at high energies. In this way,  our assumptions of fermion partial compositeness and hidden local symmetry will be encoded in the resulting form factors.
The leading SO(5) invariant terms built from the SM low energy degrees of freedom are
\begin{align}
\label{eq:Leff_f}\mathcal{L}_{eff}^F=& \overline{q_l}\slashed{p}\left(\Pi_0^L(p)+\Pi_1^L(p_1,p_2) \Sigma_i \Sigma^i \right) q_l+\overline{t_r}\slashed{p}\left(\Pi_0^R(p)+\Pi_1^R(p_1,p_2) \Sigma_i \Sigma^i \right)t_r+\\
\nonumber &\qquad\qquad\qquad\qquad+\overline{q_l}\left(M_1(p_1,p_2) \Gamma^i\Sigma_i \right)t_r +h.c.\\
\label{eq:Leff_v}
\mathcal{L}_{eff}^V=&\frac{1}{2} \mathcal{P}^{\mu\nu}\left(p^2 \Pi_0(p) \Tr(A_\mu A_\nu)+ \Pi_1(p_1,p_2) \Sigma^T A_\mu A_\nu \Sigma \right)
\end{align}
where $\Pi^{L,R}_{0,1}$, $M_1$ and $\Pi_{0,1}$ are the form factors obtained from integrating out the resonances, $\Sigma$ is defined as $\Sigma\equiv U[\pi]\cdot(\vec{0}, f)^T\,$ \footnote{Here, $\vec{0}$ is a 4 entry null vector and we set the radial mode field to zero as they do not play any role in our analysis.}, and the $\Gamma^i$ with $i=1,\cdots,5$  are the $SO(5)$ gamma matrices defined in Appendix \ref{sec:MCHM5gen} . For the fermion lagrangian, the expressions are
\begin{align}
\label{eq:ff1f}& \Pi_0^L(p)=1+\Pi_{\mathbf{4}}^L(p)=1+\frac{f^2|y_L|^2}{p^2-m_4^2}, \qquad \Pi_0^R(p)=1+\Pi_{\mathbf{1}}^L(p)=1+\frac{f^2|y_L|^2}{p^2-m_1^2} \\
& \Pi_1^L(p_1,p_2)=\Pi_{\mathbf{1}}^L(p_1)-\Pi_{\mathbf{4}}^L(p_2)=|y_L|^2 \left( \frac{1}{p_1^2-m_{1}^2} - \frac{1}{p_2^2-m_{4}^2} \right),\qquad L\leftrightarrow R \\
\label{eq:ff4f}& M(p_1,p_2)=M_{\mathbf{4}}(p_1)-M_{\mathbf{1}}(p_2)= f y_L y_R \left( \frac{m_4}{p_1^2-m_4^2}-\frac{m_1}{p_2^2-m_1^2} \right),
\end{align}
with $p_{1,2}$ being the momentum of left-handed or right-handed fermion lines respectively. To obtain the expressions above, we also assume the validity of Weinberg sum rules which set $y_{L1}=y_{L4}$ and $y_{R1}=y_{R4}$ in order to get the correct vanishing asymptotic behavior as $p_{1,2}\rightarrow \infty$.

The total momentum dependent contribution to the Yukawa coupling will come from both the fermion propagators and the $tth$ form factor vertex. In the unitary gauge the non local fermion lagrangian \eqref{eq:Leff_f} is
\begin{align}
\label{eq:Leff_f_ug}
\mathcal{L}_{eff}^F=& \, \overline{q_L}\slashed{p_1}\left(\Pi_0^L(p_1)+\Pi_1^L(p_1) \frac{S_h^2}{2} \right)q_L +\overline{t_R}\slashed{p_2}\left(\Pi_0^R(p_2)+\Pi_1^R(p_2) C_h^2 \right)t_R+\\
\nonumber &\qquad\qquad\qquad\quad +\overline{t_R}\left(M_1(p_1,p_2) \frac{S_h C_h}{\sqrt{2}} \right)q_L,
\end{align}
where we have defined $S_h=\sin\left(\frac{h+v}{f}\right)$, $C_h=\cos\left(\frac{h+v}{f}\right)$. Expanding $S_h$ and $C_h$ in powers of $h/f$ \eqref{eq:Leff_f_ug}, the unrenormalized propagators are
\begin{align}
G_L(p_1)&=\frac{i}{\slashed{p_1}\left(\Pi_0^L(p_1)+\Pi_1^L(p_1) \left<S_h^2\right>/2 \right)-M(p_1)\sqrt{\xi(1-\xi)/2}}\\
G_R(p_2)&=\frac{i}{\slashed{p_2}\left(\Pi_0^R(p_2)+\Pi_1^R(p_2) \left<C_h^2\right> \right)-M(p_2)\sqrt{\xi(1-\xi)/2}}
\end{align}
in which $M(p)=M(p_1=p,p_2=p)$. The propagator must be defined with the correct pole structure, with the pole mass and unit residue at $p^2=m_t^2$. To renormalize the propagators we factor out the kinetic form factors and introduce the top mass as the pole\footnote{ Using the relation $(\slashed{p}-M(p))^{-1}=\left(1-\frac{M(p)-m_t}{\slashed{p}-m_t}\right)^{-1}(\slashed{p}-m_t)^{-1}$.}. 
\begin{align}
\label{eq:glp1}
 G_L(p_1)&=\left(\frac{1}{\Pi_0^L(p_1)+\Pi_1^L(p_1) \left<S_h^2\right>/2 }\right)\left(\frac{1}{1-\frac{M_L(p_1)-m_t}{\slashed{p_1}-m_t}}\right)\frac{i}{\slashed{p_1}-m_t}\equiv\frac{i Z_L(p_1)}{\slashed{p_1}-m_t}\\
\label{eq:grp2}
 G_R(p_2)&=\left(\frac{1}{\Pi_0^R(p_2)+\Pi_1^R(p_1) \left<C_h^2\right> }\right)\left(\frac{1}{1-\frac{M_R(p_2)-m_t}{\slashed{p_2}-m_t}}\right)\frac{i}{\slashed{p_2}-m_t}\equiv\frac{i Z_R(p_2)}{\slashed{p_2}-m_t},
\end{align}
where $M_{L,R}$ are defined as
\begin{equation}
M_L(p_1)\equiv \frac{M(p_1)\sqrt{\xi(1-\xi)/2}}{\Pi_0^L(p_1)+\Pi_1^L(p_1) \left<S_h^2\right>/2},\quad M_R(p_2)\equiv \frac{M(p_2)\sqrt{\xi(1-\xi)/2}}{\Pi_0^R(p_2)+\Pi_1^R(p_2) \left<C_h^2\right>},
\end{equation}
and the momentum dependent residues $Z_{L,R}$ are the multiplicative expressions to the propagator defined by relations \eqref{eq:glp1} and \eqref{eq:grp2}. The pole mass and unity residues are defined when the top quark is on-shell.
\begin{align}
&\left.M_L(p_1)\right|_{p_1=m_t}=\left.M_R(p_2)\right|_{p_2=m_t}=m_t,\\
&\left.Z_L(p_1)\right|_{p_1=m_t}\,\,=\left.Z_R(p_2)\right|_{p_2=m_t}\,\,=1.
\end{align}
For the three point-function, the off shell amplitude for the top Yukawa vertex before amputation is
\begin{align}
\Gamma (p_1,p_2,p_h)&=\frac{i Z_L(p_1)}{\slashed{p_1}-m_t}\, \frac{i Z_R(p_2)}{\slashed{p_2}-m_t}\, \frac{i}{p_h^2-m_h^2}\,\kappa_\xi^{\mathbf{5}}\frac{M(p_1,p_2)\sqrt{\xi(1-\xi)}}{\sqrt{2}}\\
				    &=\frac{i}{\slashed{p_1}-m_t}\, \frac{i}{\slashed{p_2}-m_t}\, \frac{i}{(p_1-p_2)^2-m_h^2}\,y_t \kt5 f_{\tth}(p_1,p_2)
\end{align}
where $\kt5 = \frac{1-2\xi}{\sqrt{1-\xi}}$ is the misalignment suppression expected from the on-shell MCHM$_5$ and $y_t=\frac{\sqrt{2}m_t}{v}$ is the SM top Yukawa coupling. Finally, the top Yukawa form factor is given by
\begin{align}
\nonumber f_{\tth}(p_1,p_2)=&\left(1-\frac{M_L(p_1)-m_t}{\slashed{p_1}-m_t}\right)^{-1}\left(1-\frac{M_R(p_2)-m_t}{\slashed{p_2}-m_t}\right)^{-1}\\
\label{eq:tthff}&\frac{M(p_1,p_2)\left(1-2\xi\right)/\sqrt{2}}{\left(\Pi_0^L(p_1)+\Pi_1^L(p_1)\frac{1}{2} \left<{S_h}^2\right>\right)\left(\Pi_0^R(p_2)+\Pi_1^R(p_2) \left<{C_h}^2\right>\right)}.
\end{align}

Analogously to the top Yukawa form factor, the EW gauge couplings to the Higgs also receive resonance contributions from the two and three point-functions. In the unitary gauge, the vector lagrangian \eqref{eq:Leff_v} becomes
\begin{align}
\mathcal{L}_{eff}^V=&\frac{1}{2} \mathcal{P}^{\mu\nu}\left[\left(p^2 \Pi_0(p)+\frac{f^2 S_h^2}{4}\Pi_1(p_1,p_2)\right) B_\mu B_\nu\right. \\
&\nonumber \left.+ \left(p^2 \Pi_0(p)+\frac{f^2 S_h^2}{4}\Pi_1(p_1,p_2)\right)W_\mu^{a_L} W_\nu^{a_L} -2\left(\frac{f^2 S_h^2}{4}\Pi_1(p_1,p_2)\right) W_\mu^3 B_\nu\right]
\end{align} 
where $\mathcal{P}^{\mu\nu}=\eta^{\mu\nu}-\frac{p^\mu p^\nu}{p^2}$ is the transverse polarization projector, and $p_{1,2}$ are the momenta of the two gauge bosons in the Higgs interaction. The unrenormalized gauge bosons propagator can be defined after expanding $S_h^2$.
\begin{equation}
G_V^{\mu\nu}=\left(\eta^{\mu\nu}-\frac{p^\mu p^\nu}{m_V^2}\right)\frac{i Z_V(p)}{p^2-m_V^2},
\end{equation}
with $Z_V(p)$ and $M_V^2(p)$ defined as
\begin{align}
&Z_V(p)\equiv\left(\frac{1}{1-\frac{M_V^2(p)-m_V^2}{p^2-m_v^2}}\right)\frac{1}{\Pi_0(p)}\\
&M_V^2(p)\equiv-\frac{v^2}{4}\frac{\Pi_1(p)}{\Pi_0(p)},
\end{align}
satisfying the on shell conditions
\begin{equation}
\left.M_V^2(p)\right|_{p^2=m_V^2}=m_V^2, \qquad \left.Z_V(p)\right|_{p^2=m_v^2}=1.
\end{equation}
The expressions for the gauge boson form factors, obtained under the assumtion of the validity of Weinberg sum rules, are
\begin{align}
&\frac{\Pi_0(p)}{g^2}=\left(1+\frac{1}{g_\rho^2}\frac{ p^2}{p^2-m_\rho^2}\right)\\
&\frac{\Pi_1(p_1,p_2)}{g^2}=1-\frac{ p_1^2 p_2^2}{(p_1^2-m_\rho^2)(p_2^2-m_\rho^2)}
\end{align}

Thus, we compute the off shell amplitude for the Gauge-Higgs three point interaction, which results in
\begin{align}
\Gamma^{\mu\nu\alpha\beta} (p_1,p_2,p_h)&=\frac{i Z_V(p_1)\mathcal{P}_1^{\mu\alpha}}{p_1^2-m_V^2}\, \frac{i Z_V(p_2)\mathcal{P}_2^{\beta\nu}}{p_2^2-m_V^2}\, \frac{i}{p_h^2-m_h^2}\,\kappa_\xi \frac{\Pi_1(p_1,p_2)}{4}\\
				    &=\frac{i \mathcal{P}_1^{\mu\alpha}}{p_1^2-m_V^2}\, \frac{i\mathcal{P}_2^{\beta\nu}}{p_2^2-m_V^2}\, \frac{i}{(p_1-p_2)^2-m_h^2}\, g M_V \kappa_\xi f_{VVh}(p_1,p_2),
\end{align}
where $\kappa_\xi = \sqrt{1-\xi}$ is the misalignment suppression expected from the on-shell gauge coupling of the MCHM$_5$. Finally, form factor for $ZZh$ is given by
\begin{align}
\label{eq:zzhff} f_{ZZh}(p_1,p_2)=\left(1-\frac{M_Z^2(p_1)-m_Z^2}{p_1^2-m_Z^2}\right)^{-1}\left(1-\frac{M_Z^2(p_2)-m_Z^2}{p_2^2-m_Z^2}\right)^{-1} \frac{\Pi_1(p_1,p_2)}{4\Pi_0(p_1)\Pi_0(p_2)}~.
\end{align}
There is a similar expression for the $WWh$ form factor, that can be obtained by an analogous procedure. 
\begin{figure}[tbp]
\begin{center}
\includegraphics[width=0.465\textwidth]{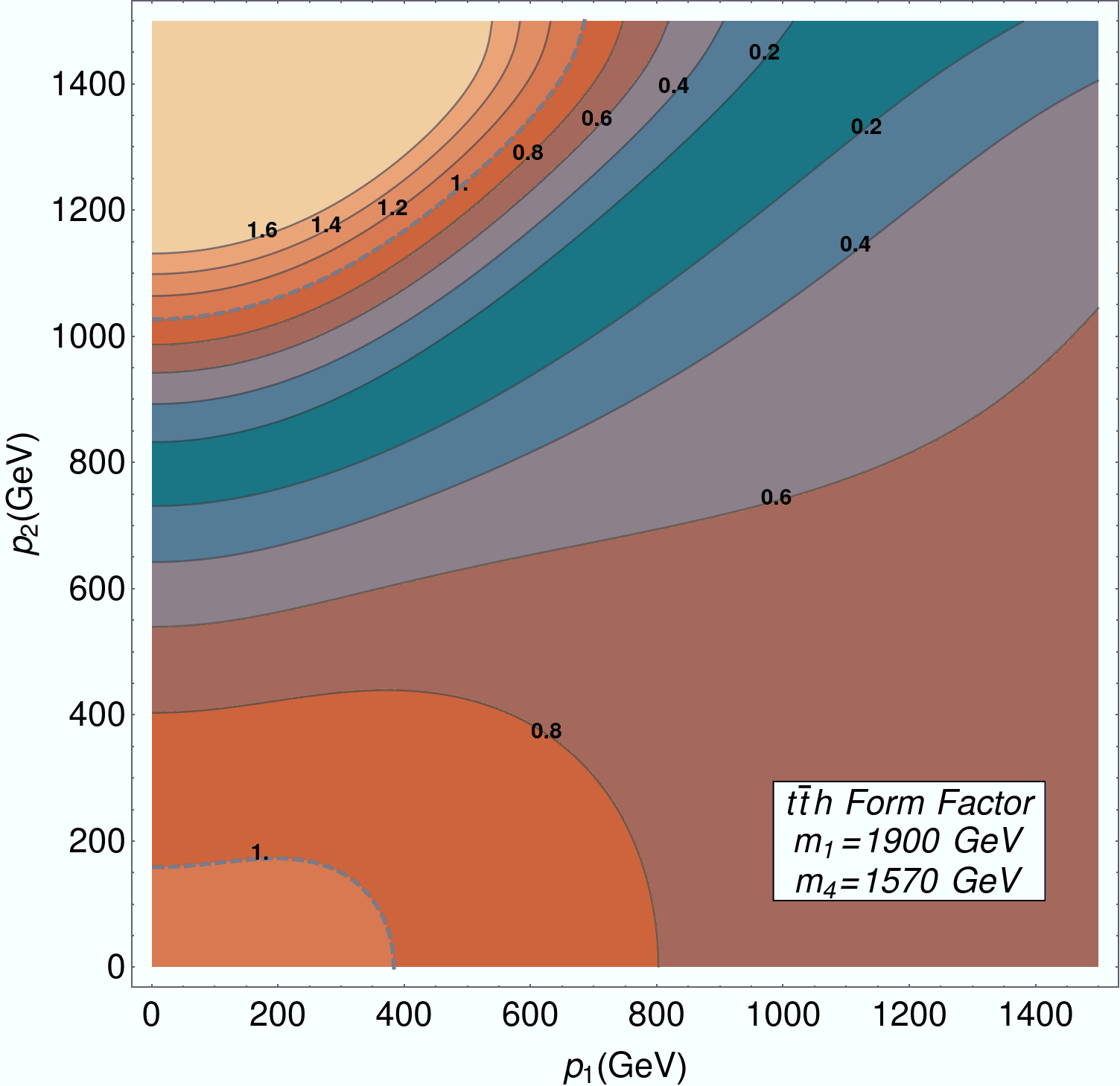} \quad \includegraphics[width=0.47\textwidth]{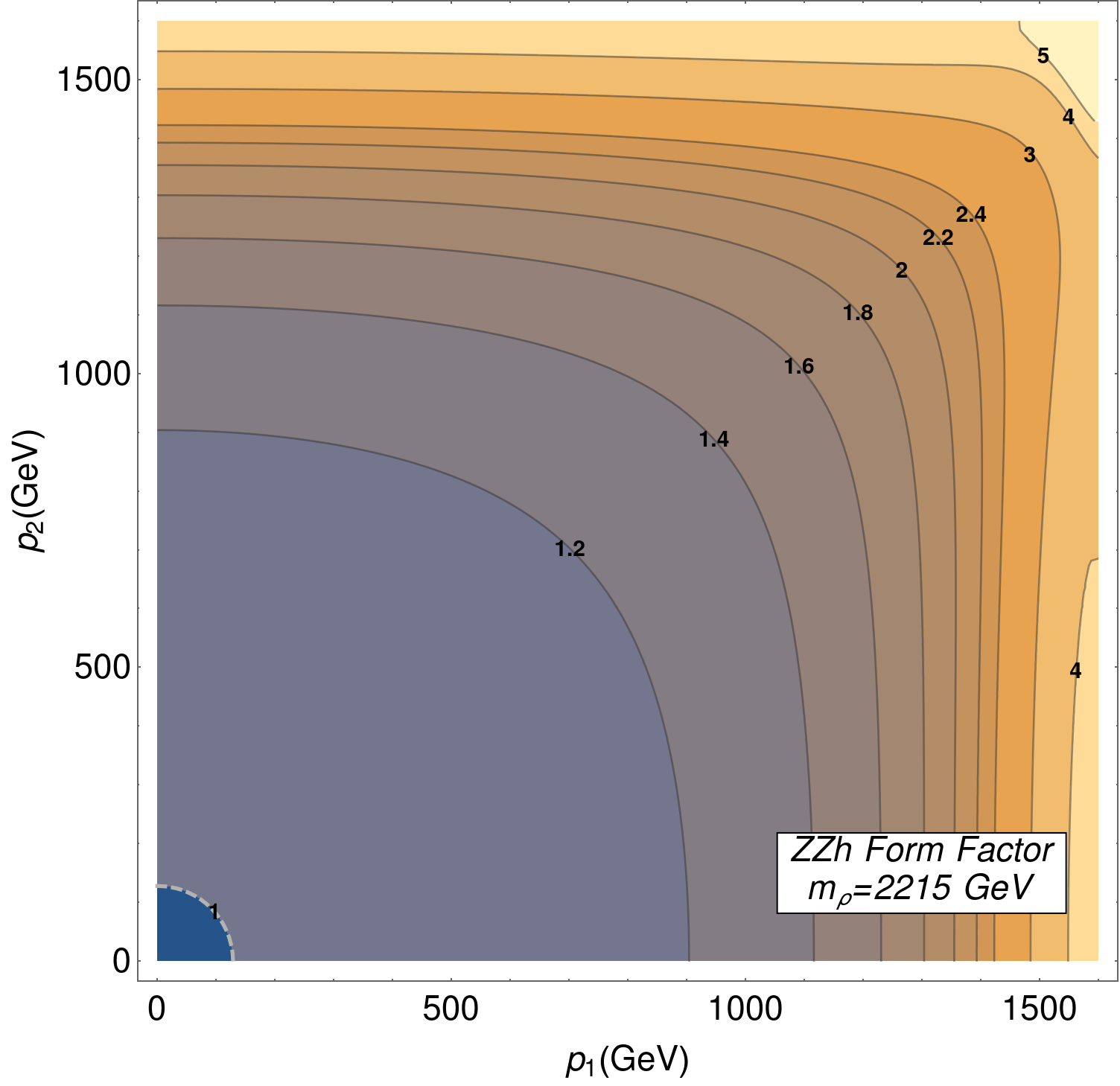}
\caption{Left: $t\bar t h$ form factor from (\ref{eq:tthff}). Right: $ZZh$ form factor from (\ref{eq:zzhff}).  }
\label{fig:ffs1}
\end{center}
\end{figure}

Both the MCHM$_\mathbf{5}$ form factors of \eqref{eq:tthff} for fermions and \eqref{eq:zzhff} for gauge boson couplings to the Higgs have similar  structure. In both cases, when the top quarks or gauge bosons are on shell, the form factor is normalized to one, and the coupling modification corresponds to  the suppression of $\kt5$ and $\kappa_\xi^V$, as obtained previously~\cite{Contino:2006qr,Panico:2015jxa}. Thus, for this example, the zero momentum transfer limit
reverts back to the predictions already known in these models. 

\begin{figure}[tbp]
\begin{center}
\includegraphics[width=0.465\textwidth]{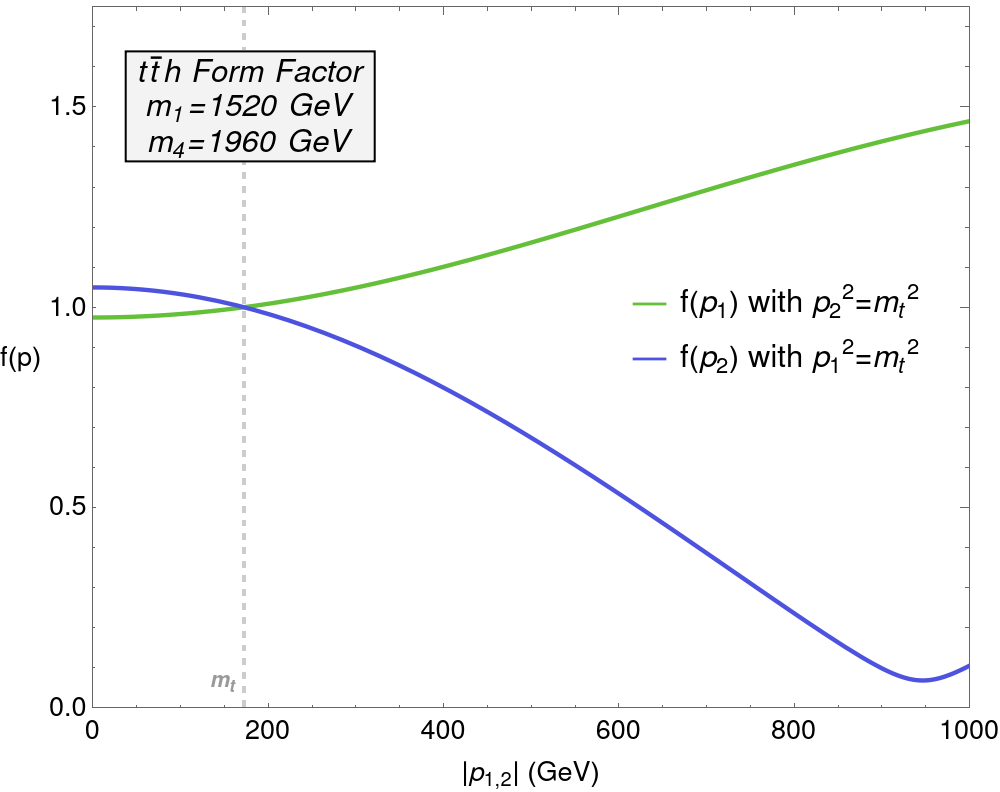} \quad \includegraphics[width=0.47\textwidth]{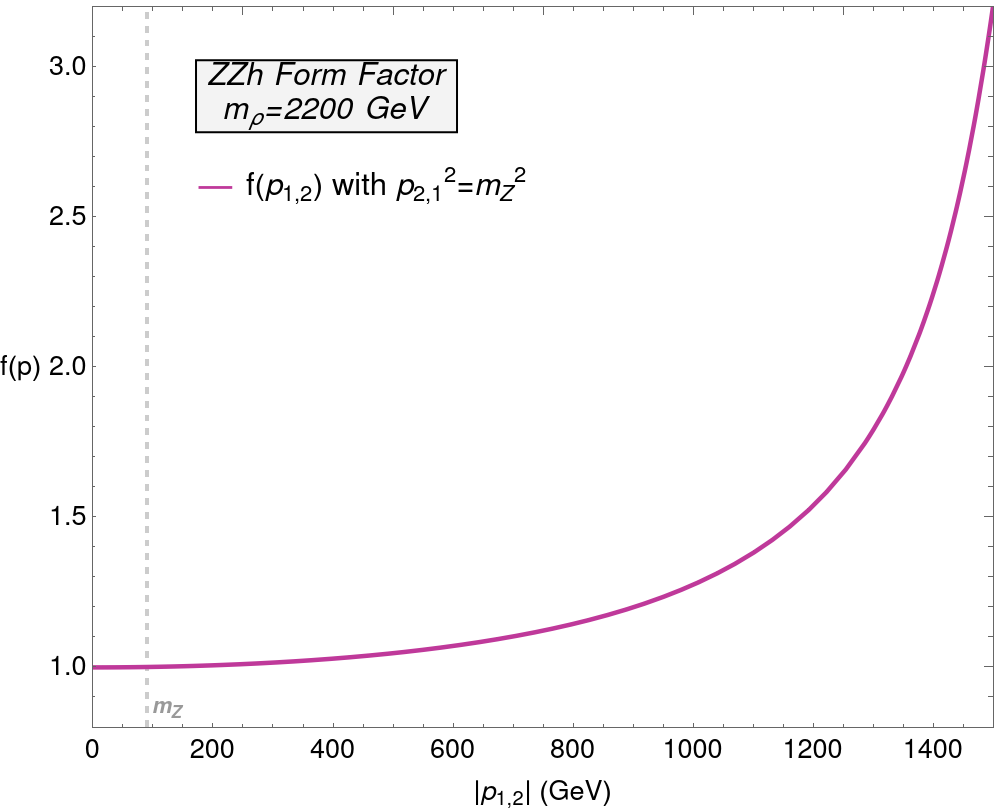}
\caption{Left:1D projections of the  $t\bar t h$ form factor from (\ref{eq:tthff}). Right: 1 D projection of the $ZZh$ form factor from (\ref{eq:zzhff}).  }
\label{fig:ffs2}
\end{center}
\end{figure}

The effects of the momentum dependence in the form factors is shown in Figure~\ref{fig:ffs1}, 
where we display the absolute value of the form factors as a function of the external momenta. When unrestricted to the on-shell contraints, large off-shell momenta in the top and gauge lines will induce important deviations from the misalignment suppression.
For illustration, we show one dimensional projections of the form factors in Figure~\ref{fig:ffs2}.  They can result in  enhancements or suppressions. 
But in both cases is clear that observing these momentum dependent effects will require channels where either the top quark or the gauge bosons are considerably 
off shell. We further explore  these issues in Section \ref{sec:coll}.

Even though we have focused on the explicit construction of the form factors within the particular case of the MCHM$_\mathbf{5}$, we expect that all CHMs will share the momentum dependent features described before. Firstly, all CHMs share the mechanism in which the composite resonances are introduced. The partial compositeness hypothesis fixes the mixing terms  that appear in the Higgs couplings to fermions. Additionally, the inclusion of vector resonances is performed in a basis independent manner. Secondly, the differences between specific implementations lie in the choice of global symmetry group and the corresponding  coset, as well as the fermion representations, both affecting the  the resonance spectrum. These may alter the form factor expressions, but the phenomenological effects of having momentum dependent couplings should be preserved. Since the form factor analytic structure corresponds to the assumed resonance spectrum, the most relevant differences between different CHMs momentum effects should come from the different possible resonant spectra.

\section{Other Examples} 
\label{sec:other}
The example of CHMs from the previous section is illustrative but by no means exhaustive. There are various other ways of generating momentum dependent Higgs couplings.  In this section we
consider two cases that will give a diversity of functional forms for the resulting form factors, which then can be used for the phenomenology of these momentum dependent effects.
This is desirable since 
for three point functions, the momentum dependence structure is rather  complex and a Källén–Lehmann decomposition is not known. So, in order to construct explicit examples, we need to assume the structure of interactions that generates the Higgs couplings to SM fermions and gauge bosons.
One lesson to be taken by the MCHM$_\mathbf{5}$ discussion of Section \ref{sec:mchm} is that elementary/composite mixings are a way to generate tree level form factors.
In what follows we first consider the direct mixing of a heavy scalar resonance with the Higgs, inducing a form factor that couples to the Higgs momentum, instead of the momenta of the fermions or gauge bosons coupled to the Higgs.
We will also consider form factors generated by a continuum, as opposed to resonances. The physics scenarios resulting in these forms of momentum dependent couplings are typically associated with quasi-conformal sectors and/or unparticle sectors. 

\subsection{Heavy Scalars}
\label{sec:heavy_scalar}
Here we consider a simple  model with one additional scalar doublet. Unlike a typical two Higgs doublet model, the second doublet does not acquire a vacuum expectation value\footnote{This also does not correspond to an inert two Higgs doublet model scenario, since we do not assume a $Z_2$ symmetry.}. This situation may arise in a variety of models. An example is the low energy theory of the model in Ref.~\cite{Hill:2019cce}. Our interest here is to present a situation where a form factor in the Higgs couplings arises where the resonance is in the Higgs momentum, unlike the situations in the couplings generated in the CHM of the previous Section.
Additionally, a  new scalar can arise in  other extensions of the electroweak theory.
The main point is that  the heavy scalar has electroweak quantum numbers so it  interacts with SM states.  Working in a scalar non diagonal basis, there is  a linear mixing between the Higgs and the new scalar. This implementation would allow the construction of form factors as we did before, as a mixing and subsequent interaction of the heavy scalars with fermions and gauge bosons will induce momentum dependent Higgs couplings.

The lagrangian describing the scalar sector of the theory with two scalar doublets 
$H_a$ and  $H_b$  is given by
\begin{align}
  \nonumber
{\cal L} =   |D_\mu H_a|^2 +|D_\mu H_b|^2&-M_a^2 H_a^\dagger H_a - M_b^2 H_b^\dagger H_b -\mu^2(H_a^\dagger H_b+ h.c.)\\
\label{eq:lagscalar}  &-\frac{\lambda}{2}(H_a^\dagger H_a + H_b^\dagger H_b)^2+\lambda' (H_a^\dagger H_b H_b^\dagger H_a) ,
\end{align}
where $D_\mu=\partial_\mu -ig W^a \frac{\sigma^a}{2}-ig' Y Z_\mu$, $H_a$ will play the role of the  lighter standard model Higgs, and we neglected a possible phase difference between the scalar doublets. The mixing term with the coefficient $\mu^2$ will induce a momentum dependence in the couplings of the lightest state, the standard model Higgs. Imposing that the lightest state has a two point function with a pole at $m_h$, 
results in the condition
\begin{align}
  \label{eq:scalarmass}&M_0^2=M_a^2 + \frac{\mu^4}{M_0^2-M_b^2}\simeq M_a^2-\frac{\mu^4}{M_b^2}
  \end{align}
where $M_0^2\simeq -(88.4 \text{ GeV})^2$, the Higgs boson mass is  $m_h=\sqrt{2}\,|M_0|$, and we used the approximation 
$|M_0^2|\ll M_b^2$.  When the lightest doublet gets a vacuum expectation value,  it induces a  small one in the heavier doublet. It is advantageous to stay in the mixed basis, where we impose the renormalization conditions for the SM Higgs, as well  as for the massive gauge bosons.
The form factor of the lightest CP even scalar, $h$,  coupling to the massive gauge bosons is given then by 
\begin{equation}
\label{eq:scalGauge}
f_{hVV}(q^2)=\frac{g_V^2 v}{2}\,\left(1-\frac{\mu^4}{M_{b}^2}\frac{1}{q^2-M_b^2}\right),
\end{equation}
where $g_V$ corresponds to the couplings obtained from (\ref{eq:lagscalar}) for $V=W^\pm, Z$, and are given by $g_{W}=g$ and $g_{Z} = \sqrt{g^2+g'^2}$ respectively.
\begin{figure}[tbp]
\begin{center}
\includegraphics[width=10cm]{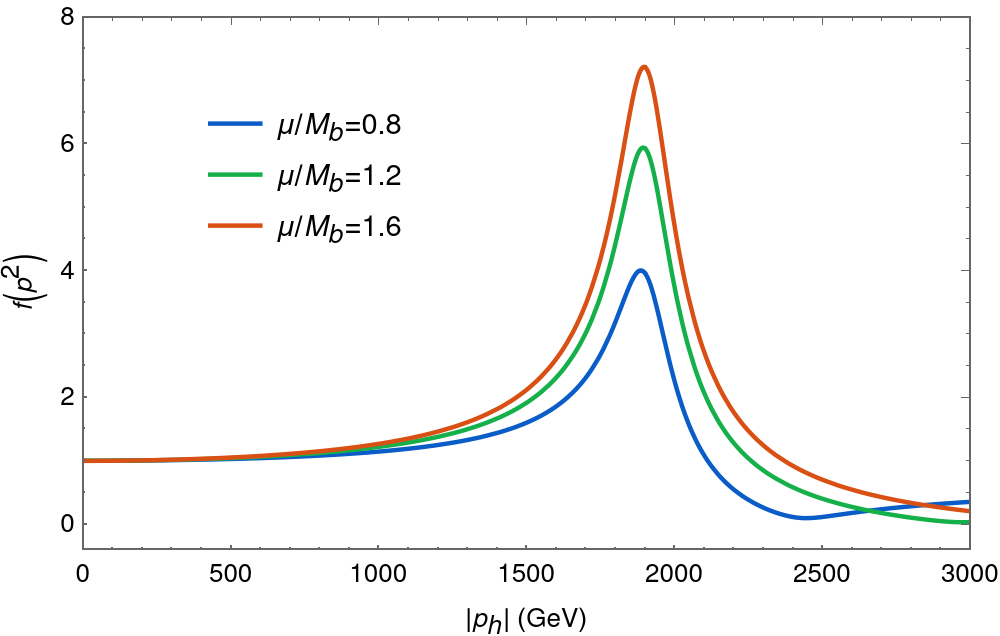} 
\caption{Higgs-Gauge Boson form factor from scalar exchange, (\ref{eq:scalGauge2}), normalized by $2M_V^2/v$, for $M_b=1.9~$TeV.  }
\label{fig:ffscalar}
\end{center}
\end{figure}
On the other hand, the gauge boson masses are defined by this coupling at $q^2=0$. 
For instance, we have
\begin{equation}
  M_W =  \frac{g\, v}{2}\,\left(1+\frac{\mu^4}{M^4_b} \right)^{1/2}~,
  \label{eq:mwmod}
  \end{equation} 
  and analogously, for the $Z$,
  \begin{equation}
  M_Z =  \frac{g_Z\, v}{2}\,\left(1+\frac{\mu^4}{M^4_b}\right)^{1/2} ~.
  \label{eq:mzmod}
\end{equation}
We then write the form factor of the Higgs boson coupled to  the massive gauge bosons in term of these masses. This is given by
\begin{equation}
\label{eq:scalGauge2}
f_{hVV}(q^2)=\frac{2 M_V^2}{v}\,\left(1-\frac{\mu^4}{M_{b}^4}\frac{q^2}{q^2-M_b^2}\right).
\end{equation}
We see that the on shell {\em couplings } of the physical Higgs boson to the $W^\pm$ and the $Z$ are modified, 
in analogy with other BSM scenarios (e.g. the MCHM studied in the previous section).
As discussed in Section \ref{sec:form1}, such on-shell coupling modification is a generic feature of SM extensions. The form factors in  \eqref{eq:scalGauge2} for the on shell Higgs result in the couplings
\begin{align}
f_{hVV}(m_h^2)&=\frac{2 M_V^2}{v}\, (1+c_b \xi)~,
                \label{eq:ffhvvonshell}
\end{align}
where we neglected corrections of order $m_h^4/M_b^4$ and higher, we defined $\xi=\frac{v^2}{M_b^2}$,  and for this case we have 
\begin{equation}
c_b = 2\lambda\,\frac{\mu^4}{M_b^4}~.
\end{equation}
The form factor is shown Figure~\ref{fig:ffscalar}, for $M_b=1.9~$TeV. We see once again here that there is a coupling modification of the order of $\xi$, here further suppressed by the ratio of the mixing $\mu$ to the mass scale of the heavier scalar. In addition, the form factor (\ref{eq:scalGauge2}) encodes the momentum dependence of the coupling modification, here driven by the presence of a scalar resonance mixing with the SM Higgs boson. In order to observe this momentum dependent effect, it will be necessary to test the off shell behavior of the
Higgs coupling in (\ref{eq:scalGauge2}). For this purpose, for instance, the study of 
$h^*\rightarrow Z^{(*)}Z^{(*)}$ proposed in \cite{Goncalves:2017iub} would be potentially sensitive to these effects.
We address the phenomenology of these form factors in Section \ref{sec:coll}.

\subsection{Continuum Form Factors}

The Higgs form factors presented so far were obtained by considering the mixing of resonances  either with the Higgs or with the SM particles in the Higgs vertices. This implies that the form factor analytic structure we probed is dominated by poles at the resonance masses for time-like momenta. Alternatively, the spectral decomposition tells us that the amplitudes from branch cuts could have a richer momentum dependence.
In particular, we  consider here  pure continuum effects arising from conformal sectors. Scale invariance is an important ingredient to arrive at this category as massive resonances are not allowed by symmetry, resulting in a model without high energy poles. Most notably, models of unparticle physics proposed by Georgi~\cite{Georgi:2007ek} account for physics without single particle states.
The interesting cases in which the unparticle sector carries gauge quantum numbers require the existence of an infrared cutoff with a mass gap as we do not want to affect the low energy part of the SM.
For instance, this is the case in Ref.~\cite{Stancato:2008mp}, where  the Higgs was proposed to be part of an approximate CFT sector with scaling dimensions larger than one. This construction allows for natural EWSB and has been studied phenomenologically~\cite{Englert:2012dq,Englert:2012cb}. Other solutions to the naturalness problem were obtained with continuum spectra. In Ref.~\cite{Bellazzini:2015cgj2}, a quantum critical point explains the lightness of the Higgs, and in Ref.~\cite{Csaki:2018kxb} naturalness is achieved by cutting the quadratic divergence with continuum top and gauge partners.

Here we make  use these continuum models in order to  find an example of form factors with branch cuts associated with physical spectral functions.
For this purpose, we consider the model of Ref.~\cite{Fox:2007sy},  of a scalar unparticle that mixes with the Higgs and derive the corresponding form factor. The scalar unparticle belongs to a broken CFT sector with an IR cutoff scale $\mu$. At this level, we remain agnostic about how the CFT is broken at low energies, simply assuming that the continuum spectral density has an energy gap at $\mu$. The scalar unparticle operator, $\mathcal{O}(x)$, of scaling dimension $d$ has the following two point function \cite{Fox:2007sy}
\begin{align}
\Delta(p,\mu,d)=\int d^4x \langle 0|\mathcal{T}\mathcal{O}(x)\mathcal{O}^\dagger(0)|0\rangle=\frac{A_d}{2\pi}\int_{\mu^2}^\infty ds (s-\mu^2)^{d-2}\frac{i}{p^2-s+i\epsilon},
\label{eq:ScUnp}
\end{align}
where $A_d=\tfrac{16\pi^{5/2}}{(2\pi)^d}\tfrac{\Gamma(d+1/2)}{\Gamma(d-1)\Gamma(2d)}$ is a chosen normalization for the phase space element of the scalar unparticle. The spectral function, defined for $1\leq d < 2$, is given by
\begin{equation}
\rho(s) =
	\begin{cases}
		\frac{A_d}{2\pi} (s-\mu^2)^{d-2} \qquad \text{for } s>\mu^2, \\
		0 \qquad \text{otherwise.}
	\end{cases}
\end{equation}
With such ansatz for the spectral density, we can integrate the spectral decomposition to obtain
\begin{equation}
\Delta(p,\mu,d)=\frac{A_d}{2\sin{d\pi}}\frac{i}{(\mu^2-p^2-i\epsilon)^{2-d}},
\label{eq:unp_prop}
\end{equation}
The following effective action can be used to derive such propagator,
\begin{align}
S_{\rm NL}=\int \frac{d^4 p}{(2\pi)^4} \phi^\dagger\left[p^2-\mu^2\right]^{2-d}\phi(p)\equiv\int \frac{d^4 p}{(2\pi)^4} \phi^\dagger\Sigma(p^2)\phi(p).
\end{align}
Now, if we assume that the scalar unparticle is charged just like the Higgs with SM quantum numbers,  the non local unparticle scalar action is similar to the Unhiggs case of Ref.~\cite{Stancato:2008mp}
\begin{equation}
  S_{\rm NL}=\int d^4x \left\{ \phi^\dagger (D^2-\mu^2)^{2-d} \phi -\lambda_t \overline{u}_R\frac{\phi^\dagger}{\Lambda^{d-1}}q_L+h.c. +\mathcal{L}_{mix}\right\},
  \label{nonlocal1}
\end{equation}
where $D_\mu$ is the electroweak covariant derivative and we added a mixing term to allow for the generation of an unparticle-Higgs form factor,
\begin{equation}
\mathcal{L}_{mix}=\alpha|H|^2\frac{|\phi|^2}{\Lambda^{2(d-1)}}.
\end{equation}

\begin{figure}
\centering
\includegraphics[scale=0.1]{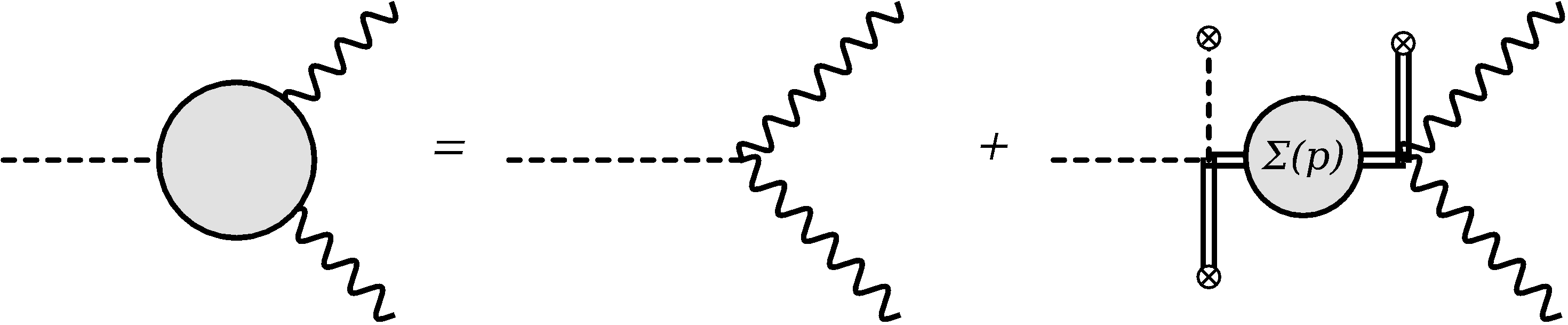}
\caption{Higgs form factor from mixing with an unparticle scalar sector.}
\label{fig:cont_ff}
\end{figure}

\begin{figure}[tbp]
\begin{center}
\includegraphics[width=0.465\textwidth]{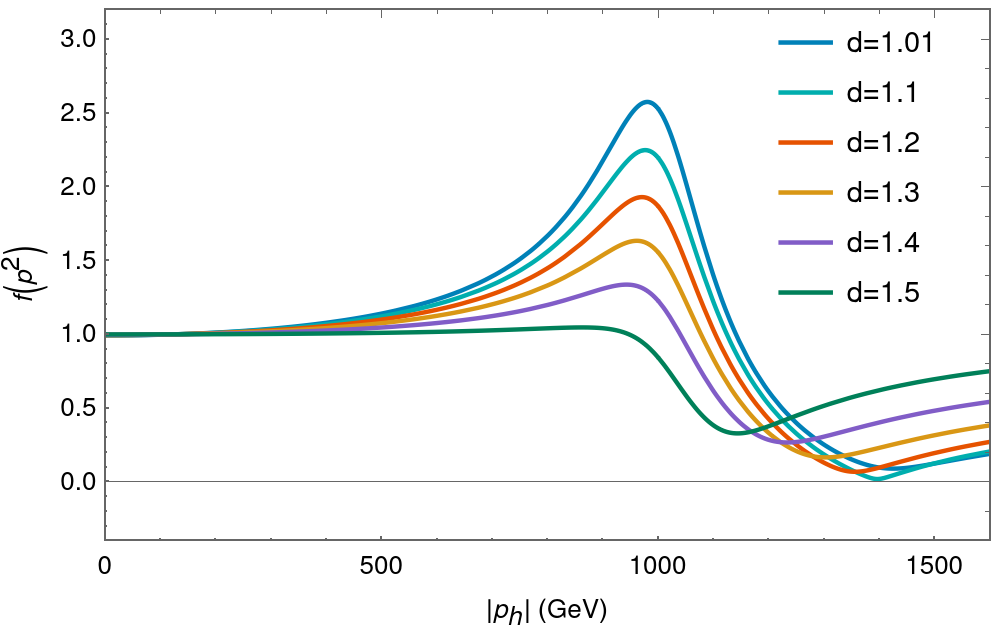} \quad \includegraphics[width=0.47\textwidth]{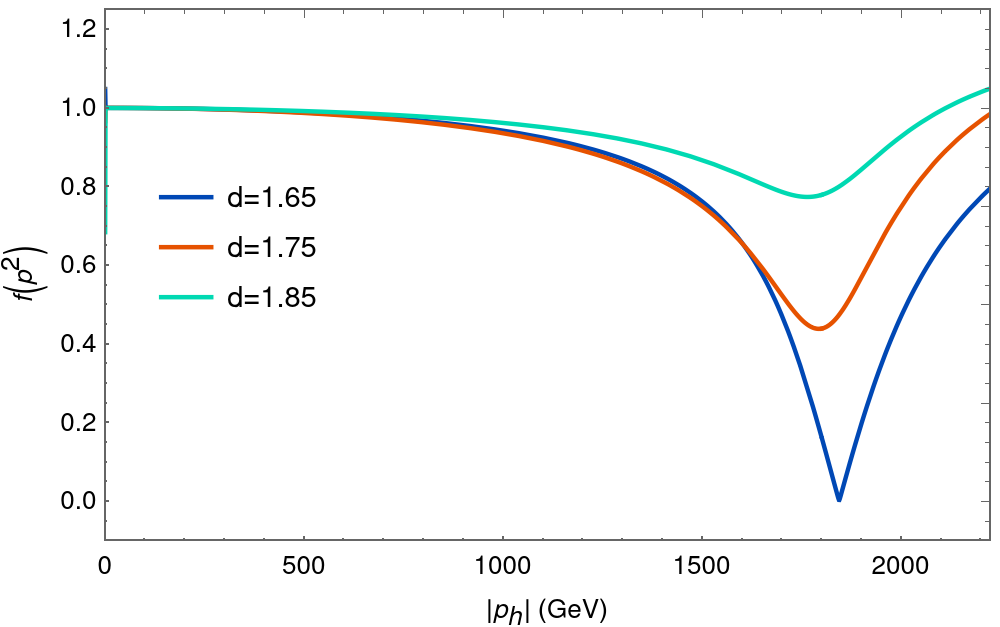}
\caption{The Higgs coupling form factor with gauge bosons, (\ref{eq:contff}), for various values of $d$, the dimension of the scalar unparticle operator, and for a $10\%$ width of the unscalar.}
\label{fig:cont_ff_d}
\end{center}
\end{figure}

Here we focus on the modifications on the interactions of the Higgs with gauge bosons. In order to obtain the effects of the unparticle sector mixing we must realize the gauge symmetry in the non-local effective action. This is done by using Mandelstam's method~\cite{Mandelstam:1962mi}, which introduces a Wilson line between two unparticle fields at different positions~\cite{Terning:1991yt}. For more details on the application of the method to the unparticle setting we refer to Refs.~\cite{Stancato:2008mp,Cacciapaglia:2008ns}.
The non local modification of the Higgs coupling to gauge bosons is illustrated by the third  diagram in Figure~\ref{fig:cont_ff}. To obtain this, we need to
take the fourth functional derivative of the non local action. When using Mandelstam's method, we do not use the action in (\ref{nonlocal1}), but the non gauged version supplemented by the appropriate Wilson line. This results in
\begin{align}
ig^2 \Gamma^{ab\alpha\beta}(p,q_1,q_2) &= \frac{\delta^4 S_{\rm NL}}{\delta A^{a\alpha}(q_1)\delta A^{b\beta}(q_2)\delta \phi^\dagger(p+q_1+q_2)\delta\phi(p)}\\
&ig^2  \left\{ \phantom{\frac{1}{2}}\left( T^a T^b + T^b T^a \right) g^{\alpha\beta}\mathcal{F} (p,q_1+q_2)  \right. \\
&\left.+ T^a T^b \frac{(2 p+q_2)^\beta (2 p + 2 q_2 + q_1)^\alpha}{q_1^2 + 2 (p+q_2)\cdot q_1} \left[ \mathcal{F} (p,q_1+q_2) - \mathcal{F} (p,q_2) \right]  \right.\nonumber \\
&\left.+ T^b T^a \frac{(2 p+q_1)^\alpha (2 p + 2 q_1 + q_2)^\beta}{q_2^2 + 2 (p+q_1)\cdot q_2} \left[ \mathcal{F} (p,q_1+q_2) - \mathcal{F} (p,q_1) \right]\right\}\,,\nonumber
\end{align}
The function $\mathcal{F}$ is defined as
\begin{equation}
\mathcal{F} (p,q) = -\,\frac{\left(\mu^2 - (p+q)^2\right)^{2-d}-\left(\mu^2-p^2\right)^{2-d}}{q^2 + 2 p\cdot q}\,.
\end{equation}
Since we are interested  in the off shell behavior of the Higgs, we can assume that the vector bosons are on shell. For instance, the amplitude for two external $Z$ bosons simplifies to
\begin{equation}
\frac{ig}{c_w}\epsilon_\alpha (q_1)\epsilon_\beta (q_2)\Gamma^{\alpha\beta}(q_1,q_2)=i\frac{g^2f^d}{2 c_w}\epsilon(q_1)\cdot\epsilon(q_2)\mathcal{F}(0,q_1+q_2)~,
\end{equation}
with analogous results for the $W^\pm$ case. 
Finally, the form factor is given by
\begin{equation}
  f_{unp}(p,q_1,q_2)=\frac{g^2v}{2c_w^2}+\frac{\alpha v f^{2d}}{\Lambda^{2(d-1)}}\frac{g_*^2}{2}\frac{A_d}{2\sin d\pi}\frac{i\mathcal{F}(0,q_1+q_2)}{(p_h^2-\mu^2)^{2-d}},
  \label{eq:contff}
\end{equation}
which follows the structure of Figure~\ref{fig:cont_ff}. Here, $f$ and $g_*$ are the VEV of the unparticle scalar its EW gauge coupling respectively. We plot the form factor above in Figure~\ref{fig:cont_ff_d}, for various values of the dimension of the scalar unparticle operator, $d$. Since the unscalar is unstable, it has a width associated to its decays which we take as a small complex shift in the branch point position $\mu^2$ in the unparticle two-point function \eqref{eq:unp_prop}.   
We see that the effect of the form factor {\em in the time-like region}  is an enhancement for smaller values of $d$, whereas for larger values of $d$ there is a suppression. 
These effects wil be further discussed  in  
Section~\ref{sec:coll}.

\section{Phenomenology at the LHC }
\label{sec:coll}

In this section, we consider form factor signals at the LHC. As discussed in the previous sections,  form factors are normalized to unity for on-shell momenta. This normalization will result in  no momentum dependent modifications if the particles from the Higgs vertices are on-shell. So, to observe the momentum dependent effects requires a significant degree of off shellness in the Higgs production channels at high energies. Which  particle  must be off shell depends on which momenta the form factor is sensitive to. For instance, for the CHM form factors we need off shell top quarks or Ws or Zs,  since the form factors are functions of $p_t^2$ and $p_Z^2$. On the other hand, for the $p_h^2$-dependent scalar and unscalar form factors, we require an off shell Higgs for  the momentum effects to be accessible. Large off shell momenta allow us to improve the reach for new physics as we extend the theory beyond the Electroweak scale. So, for momenta satisfying $p^2\gg v^2$, we expect an improvement  over expected new physics contribution of order $v^2/\Lambda^2$. We find that the form factors are useful in situations in which new states are not yet produced, are very hard to see, but their effects can be observed in the tails of the momentum distributions. Furthermore, there are cases where there are no new resonance, such as is the case with the unparticle scalar sector introduce in the previous section. 

Here, we obtain the parton level modifications of the SM signal induced by the form factor models presented previously. Our goal is  to show how much the signals are modified due to the Higgs momentum dependent couplings of sections \ref{sec:mchm} and \ref{sec:other}. A complete analysis of the experimental reach of the effects will be left for future studies. The implementation of the models is done using  {\tt MadGraph5\_aMC@NLO}~\cite{Alwall:2014hca}. The channels of Figure~\ref{fig:channels} are simulated with a center of mass energy of $\sqrt{s}=14\text{ TeV}$. 

\subsection*{Composite Higgs Form Factors}

Collider searches for composite Higgs models are traditionally based on both the coupling modifications of the Higgs couplings, and on the direct searches for heavy resonances. In the MCHM$_\mathbf{5}$, the modifications to the Higgs couplings are the misalignment suppression factors of order $\xi=v^2/f^2$.
\begin{equation}
\label{eq:kappas}
\kappa_{\xi}^{V} = \sqrt{1-\xi},\qquad\qquad \kt5 = \frac{1-2\xi}{\sqrt{1-\xi}}.
\end{equation}
Current constraints from LHC Run~2 put a lower bound on the compositeness scale, $f$, to be around $780$ GeV~\cite{Khosa:2021wsu}. Direct searches for the production of heavy top partners are sensitive up to about $1.3$ TeV masses~\cite{ATLAS:2018tnt,CMS:2019eqb}, and for heavy vector resonances, the bounds are in the range of $(2-4)$ TeV~\cite{CMS:2022tdo,ATLAS:2020fry}. 

\begin{figure}[tbp]
\centering
\includegraphics[width=0.55\textwidth]{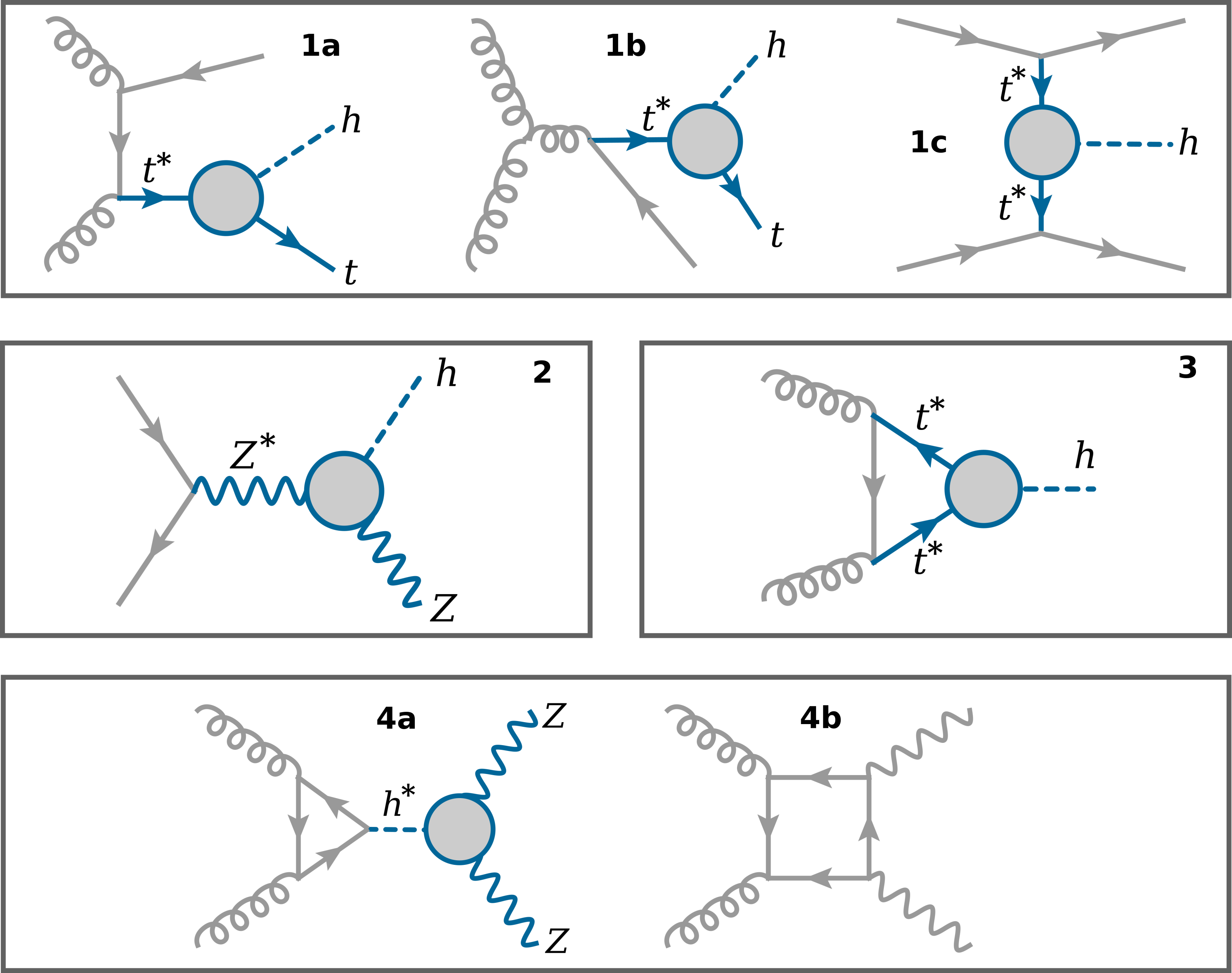}
\caption{Relevant channels for the appearance of momentum effects through the form factor vertices. The $t\overline{t}h$ channel (1) presents an off-shell top quark and the $Zh$ (2) an off-shell Z-boson. The gluon fusion (3) has off-shell top quarks in the loops and the ZZ channel (4) an off-shell higgs.  }
\label{fig:channels}
\end{figure}

The introduction of momentum dependence will alter the predictions of \eqref{eq:kappas} for off shell momenta. Since in this case the form factors are induced by mixing with either fermion or vector resonances, they will be a function of the top quark or Z boson momentum. Therefore, to achieve a $p^2/f^2$ sensitivity  greater than the $v^2/f^2$  in the on shell suppression, we should choose channels with a high degree of top or Z, W off shellness. The most important channels that have the requirements are $t\bar th$, and $Zh$ represented in Figure~\ref{fig:channels} (1a-1c), (2). 

The resonance masses used in the form factor expressions respect bounds from direct production, as well as  theoretical constraints.
Both form factors from Section \ref{sec:mchm} must respect the on shell normalization conditions, i.e. we recover the on shell suppressions  (\ref{eq:kappas}) . Consequently, if all external legs were on shell, we would only have the MCHM$_\mathbf{5}$ coupling suppression factors. 

The Higgs boson acquires a potential at one loop level due to  the explicit breaking of the $SO(5)$ by the gauging of the EW group and the Yukawa couplings. Consequently, we imposed the additional constraint of the Coleman-Weinberg calculation to ensure the obtention of a light Higgs boson mass. The Higgs mass is dominated by the top partner loops leading to~\cite{Pomarol:2012qf}
\begin{align}
\nonumber&\label{mh}m_h^2=\frac{N}{\pi^2}\left[\frac{m_t^2 \left(m_1^2 m_4^2\right) \log \left(\frac{m_1^2}{m_4^2}\right)}{f^2 \left(m_1^2-m_4^2\right)}+\frac{v^2}{4} \left(1-\frac{v^2}{f^2}\right) (\Delta y^2)^2 \left(\frac{\left(m_1^2+m_4^2\right) \log \left(\frac{m_1^2}{m_4^2}\right)}{2 \left(m_1^2-m_4^2\right)}-1\right)\right],\\
&\Delta y^2 \equiv \left| y_l\right| ^2-2 \left| y_r\right| ^2.
\end{align}
The light Higgs condition correlates the two masses $m_1$ and $m_4$ so that they cannot be both arbitrarily heavy for a fixed $f$. In general, we expect lighter fermionic resonances in CHMs to obtain a light Higgs mass~\cite{Contino:2006qr,Pomarol:2012qf}. 

\begin{figure}[tbp]
\centering
\includegraphics[width=0.65\textwidth]{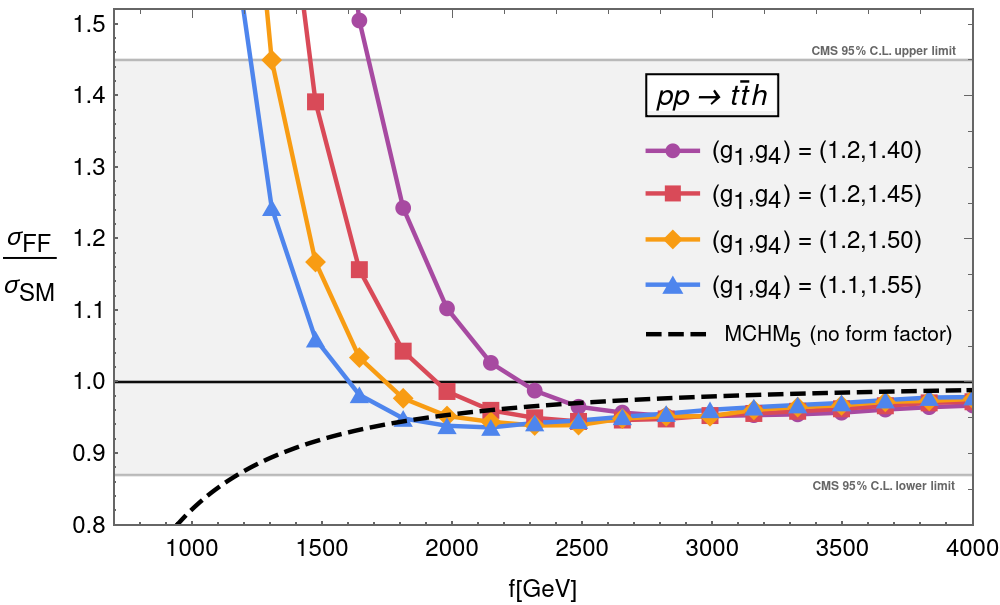}
\caption{\label{fig:cs_tth} Normalized $t\overline{t}h$ cross-section of the top-Yukawa form factor model as a function of $f$ for the MCHM$_\mathbf{5}$. Parton-level signal simulation done in {\tt MadGraph5\_aMC@NLO} at $\sqrt{s}=14$ TeV. The plot is normalized to the SM prediction. The CHM without momentum dependence has an overall suppression of vertices that goes with $\kt5$. The dashed black curve depicts such suppression. By turning on the form factor effects, we get an enhancement due to the analytic structure of their expressions. Fixing the parameters by theory constraints leaves us with a single mass dependence, which is varied to produce the different curves shown. All curves respect the bounds from direct resonance searches, and the gray band is the CMS $95\%$ confidence level bound on the experimental cross-section over SM prediction observable.
}
\end{figure}

\begin{figure}[tbp]
\begin{center}
\includegraphics[width=0.65\textwidth]{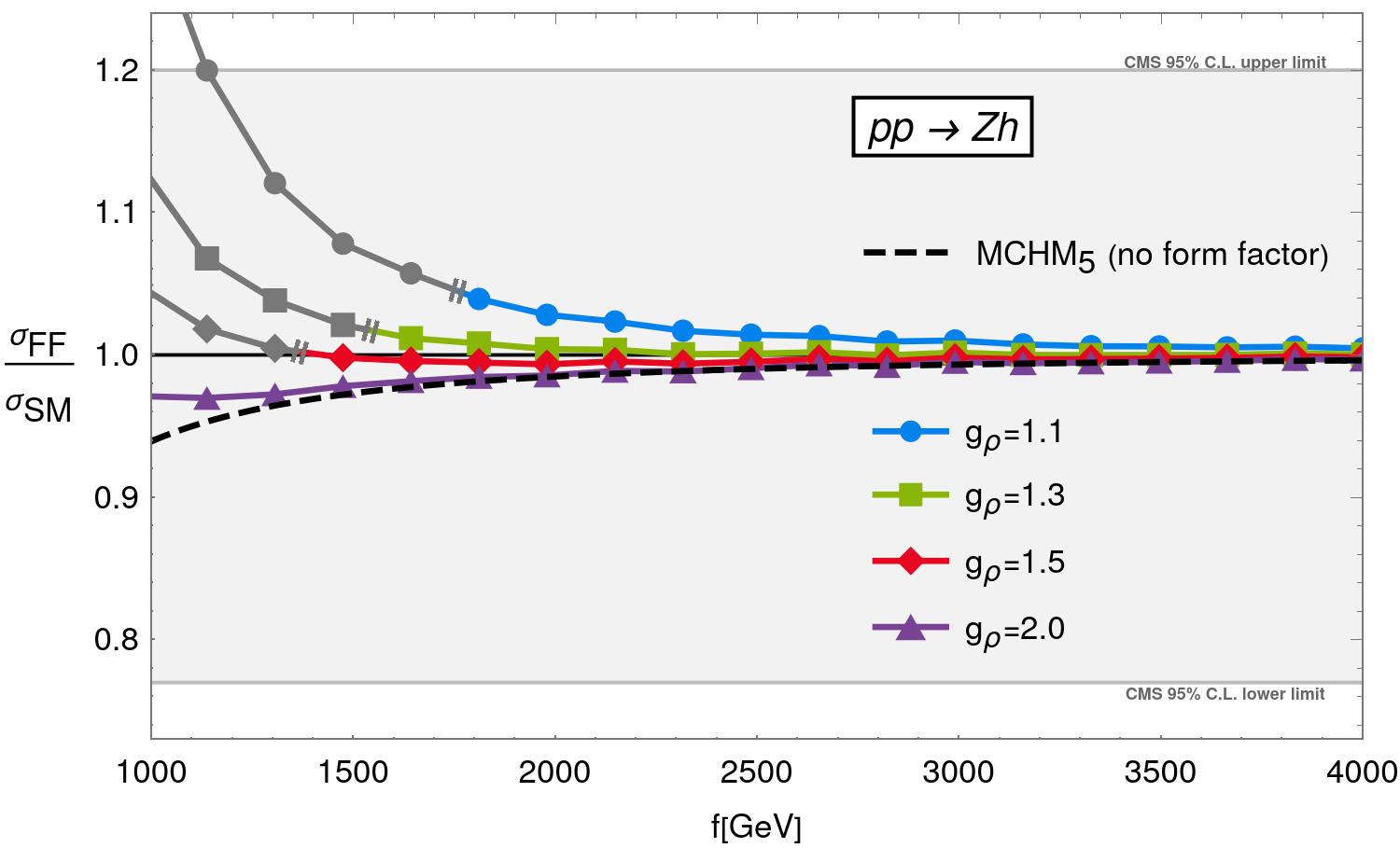}
\caption{\label{fig:cs_zh} Normalized $Zh$ cross-section of the ZZh form factor model as a function of $f$ for the MCHM$_\mathbf{5}$. Parton-level signal simulation done in {\tt MadGraph5\_aMC@NLO} at $\sqrt{s}=14$ TeV. The plot is normalized to the SM prediction. The CHM without momentum dependence has an overall suppression of vertices with $\kappa_V$. The dashed black curve depicts such suppression. By turning on the form factor effects, we get an enhancement due to the analytic structure of their expressions. The colored curves respect the bounds from direct resonance searches, and the gray curve points above the double lines are excluded by the direct searches for vectorial resonances. The gray band is the CMS $95\%$ confidence level bound on the experimental cross-section over SM prediction.
} 
\end{center}
\end{figure}

After imposing the necessary constraints, we obtain the form factor modifications to the corresponding SM cross-sections as a function of the compositeness scale $f$. The results are shown in Figures \ref{fig:cs_tth} and \ref{fig:cs_zh}. The resonance masses scale linearly with $f$, and each curve is obtained by fixing the $g$ parameter, that is, the mass to $f$ ratio. For the $t\overline{t}h$ channel, we have fixed the mass ratios of the two resonances and scaled them as $m_{1,4}\sim g f$. The ZZh vertex has only one resonance in the form factor, so we plot the different values of the $\rho$ mass for a fixed $g=m_{\rho}/f$. We assume that the width of the resonances is $10\%$ of the masses throughout all the MCHM$_\mathbf{5}$ simulations.

The form factor couplings modify the on shell low energy predictions of CHMs significantly. The momentum dependent effects result in a significant signal enhancement for lower values of $f$, corresponding to lower resonance masses for which the pole begins to be accesible at high enough off shellness. This effect is more pronounced for the top Yukawa form factor, with a single pole structure, as shown before in Figure~\ref{fig:ffs1}. On the other hand, the $ZZh$ form factor,  having  a double pole structure that falls faster for lower momenta, presents a smaller enhancement. Because of the stronger bounds on vector resonances masses, the signal enhancement is limited to about $5\%$ of the MCHM$_\mathbf{5}$ prediction in the latter case. The gray bands in the figure are the CMS $95\%$ confidence levels of the signal strength observable obtained for the LHC Run 2~\cite{CMS:2019lcn}, used here just for illustration, since in the simulation we used a center of mass energy of $14$~TeV, slightly higher than the ine in Run 2.

\begin{figure}[tbp]
\quad\,\,\includegraphics[width=0.444\textwidth]{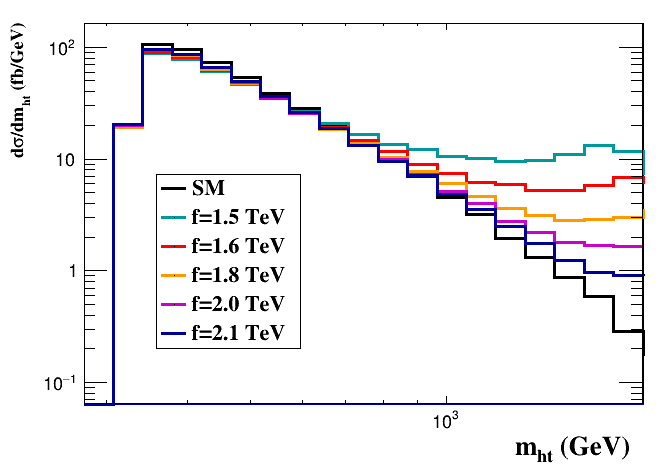} \qquad\, \includegraphics[width=0.444\textwidth]{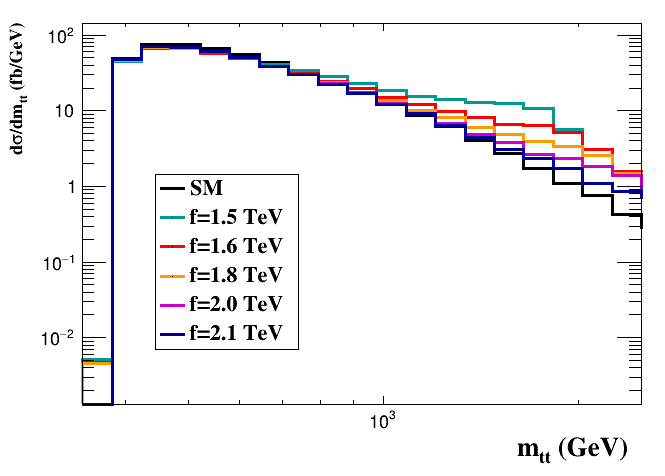}\\
\includegraphics[width=0.48\textwidth]{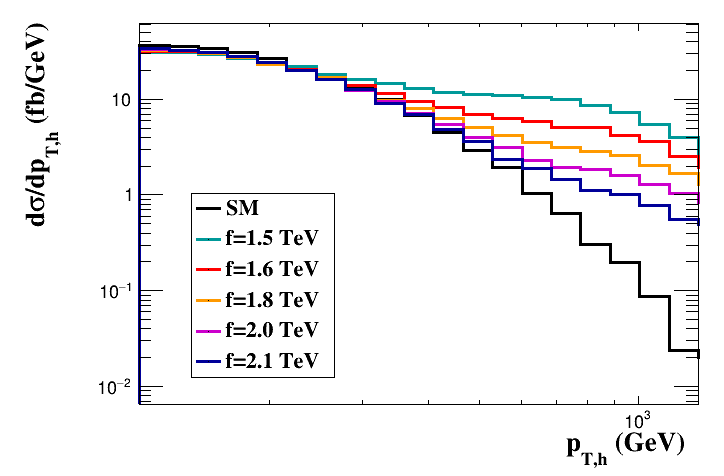} \quad \includegraphics[width=0.48\textwidth]{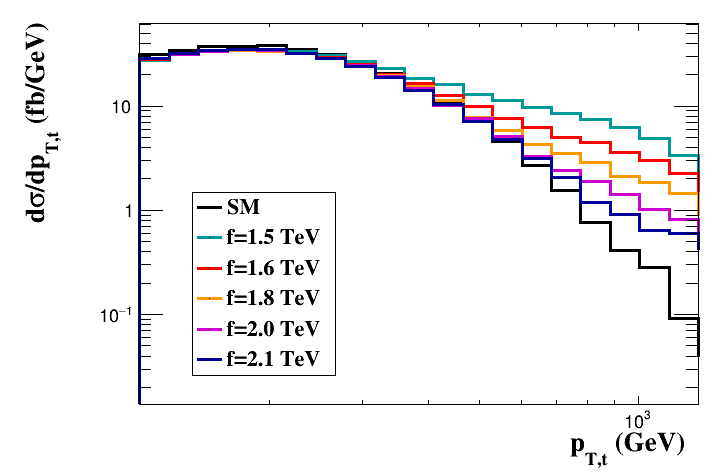}
\caption{Invariant mass and transverse momentum distributions for the $t\overline{t}h$ process with MCHM$_\mathbf{5}$ form factors. Signal simulation done in {\tt MadGraph5\_aMC@NLO} and kinematic analysis in {\tt MadAnalysis5}~\cite{Conte:2012fm}. The black curve is the SM prediction and the colored ones are the form factors for different values of the compositeness scale. The ratio of mass to $f$ is fixed as $g_1=1.1$ and $g_4=1.55$ for the MCHM$_\mathbf{5}$ fermionic resonances that enter the form factor expression.}
\label{fig:dist_mchm} 
\end{figure}

In addition to the total cross sections, the form factors also modify the shape of the kinematic distributions. In the MCHM$_\mathbf{5}$ without form factors, the coupling modifications suppress the signal only up to an overall normalization.  After including the  form factor effects, the number of events on the tails of the distributions is larger than the SM prediction, improving the prospects for precision measurements. The effect is illustrated in Figure \ref{fig:dist_mchm}. The sensitivity and reach of collider probes to form factors will be addressed in more detail  in future studies, including the analysis of background effects. We expect that the modification of the shape of distributions will be particularly important in the context of the HL-LHC, as well as in future accelerators probing the Higgs couplings with increasingly higher precision. As indicated in the cross section results, there is an intermediary range of $f$ values for which the combination of the misalignment suppression and the momentum enhancement cancel mimicking  the SM cross section prediction. Because of this effect, the kinematic distributions will be  important to determine the presence of deviations from the SM. 

To conclude the discussion regarding CHM phenomenology, we point out that the Higgs-top  form factor affects the prediction of Higgs production  through gluon fusion.
Similarly, the $W$ Higgs form factor affects the $h\to\gamma\gamma$ rate. The overall effect in $gg \to h$ is a suppression.  Here, even if the Higgs is on shell, the loop states are not.
A quick estimate shows that the effect can be as large as $10\%$ in the $gg \to h$ cross section, for a mass scale of about $1.5-2.0$~TeV, which is still allowed by the current data~\cite{ATLAS:2020qdt}. We leave a more detailed study of the resulting constraints for future work. This would include all the effects both in production and decays of the Higgs boson. 

\subsection*{Form Factors from Scalars and Unscalars}
In Section \ref{sec:other}, we presented form factors that were induced by the Higgs mixing with a new scalar particle, or an unparticle scalar sector. As we need the Higgs to be off shell to result in momentum dependence in te couplings, these form factors have a different collider phenomenology than the ones from CHMs. The most promising channels in these cases are $pp\rightarrow h^{*}\rightarrow Z^{(*)}Z^{(*)}$ with the $Z$s going to either $4l$ or $2l2\nu$ final states. Even though an off shell Higgs is unlikely to be seen in vector boson fusion at the LHC, future accelerators could be sensitive to it ~\cite{Goncalves:2017iub,Goncalves:2018pkt,Goncalves:2020vyn}. We focus on the $pp\rightarrow 4l$ channel for this work since we aim to illustrate the form factor signal effect. 

As shown in Figure~\ref{fig:channels}(4b), the $pp\rightarrow 4l$ channel has an irreducible background that leads to an interference between the Higgs signal and the box diagram. We can separete the contributions to the cross section as 
\begin{equation}
\frac{d\sigma}{dm_{4l}}=\frac{d\sigma_{h^*}}{dm_{4l}}+\frac{d\sigma_{int}}{dm_{4l}}+\frac{d\sigma_{box}}{dm_{4l}},
\label{eq:csectppzz}
\end{equation}
where $\sigma_{h^{*}}$ is the off shell Higgs contribution, $\sigma_{box}$ is the box diagram and $\sigma_{int}$ is the interference between the two. The interference is destructive due to the different relative sign between the signal and box contributions, and is larger in magnitude than the signal at the higher end of the $m_{4l}$ tail. In the SM, the signal and box have a logarithmic momentum dependence that exactly cancels and unitarizes the cross section. Thus, any modifications of the Higgs coupling at high energies will affect the balance between these contributions in \eqref{eq:csectppzz}.  

After introducing the form factor, $F({p_h})$, the cross section scale as
\begin{equation}
\frac{d\sigma}{dm_{4l}}=|F(m_{4l})|^2\frac{d\sigma_{h^*}}{dm_{4l}}+|F(m_{4l})|\frac{d\sigma_{int}}{dm_{4l}}+\frac{d\sigma_{box}}{dm_{4l}}.
\label{eq:csectppzz_withff}
\end{equation}
From \eqref{eq:csectppzz_withff}, we can see that there are various situations in which the signal is enhance with respect to the SM prediction. For $F(p^2)>1$ at the tail of the distribution, the total number of events should increase above the SM expectation if the quadratic form factor term can overcome the linear interference contribution. Conversely, for $F(p^2)<1$ at high energies, the cross-section can also be enhanced compared to the SM if the box contribution becomes dominant over the interference and signal. So, even for a suppressed Higgs coupling, we can get an enhanced differential cross section.

\begin{figure}[tbp]
\begin{center}
\includegraphics[width=0.45\textwidth]{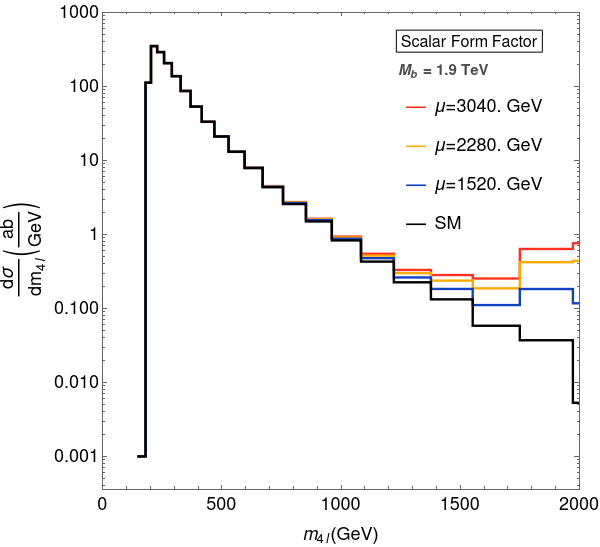} \qquad \includegraphics[width=0.43\textwidth]{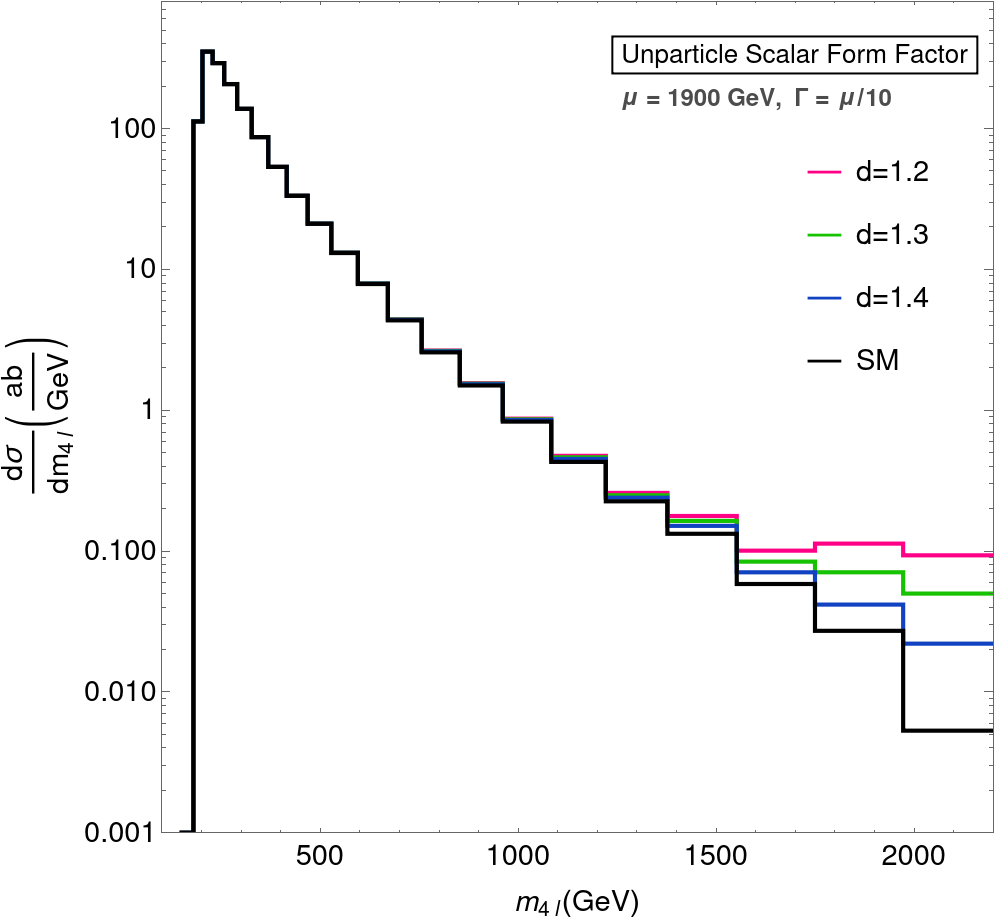}\\
 \includegraphics[width=0.44\textwidth]{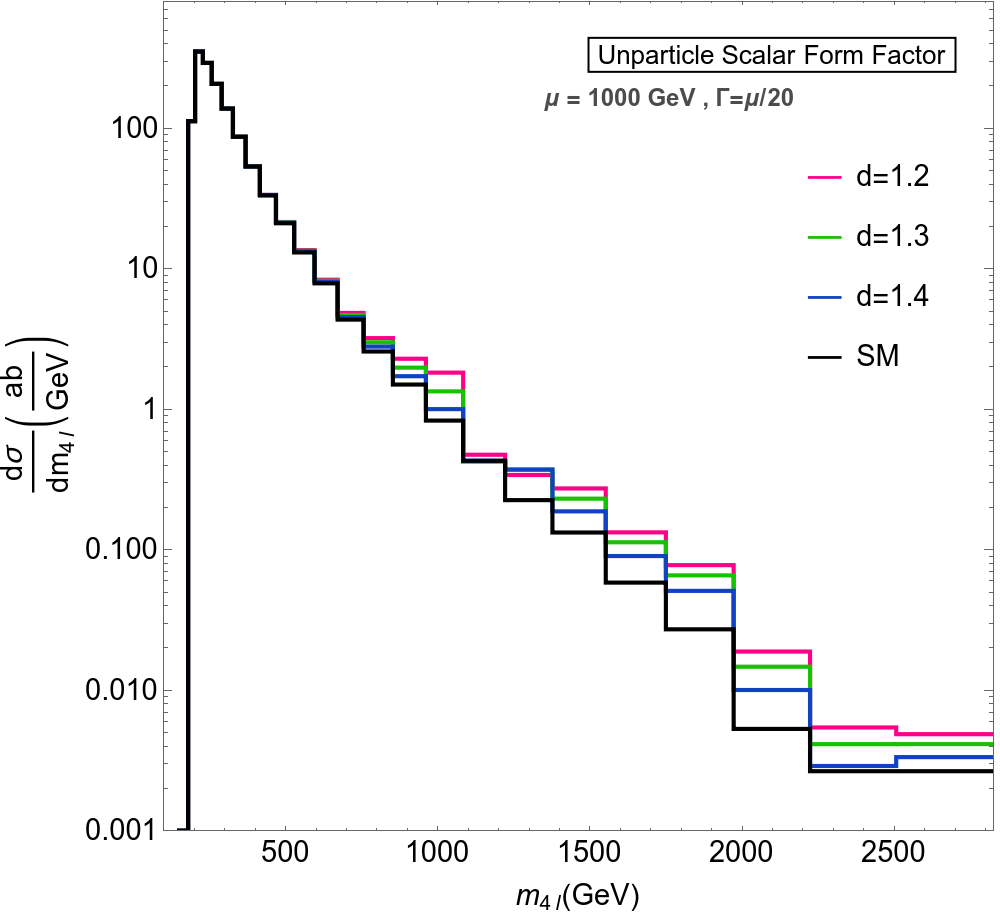} \qquad\,\, \includegraphics[width=0.435\textwidth]{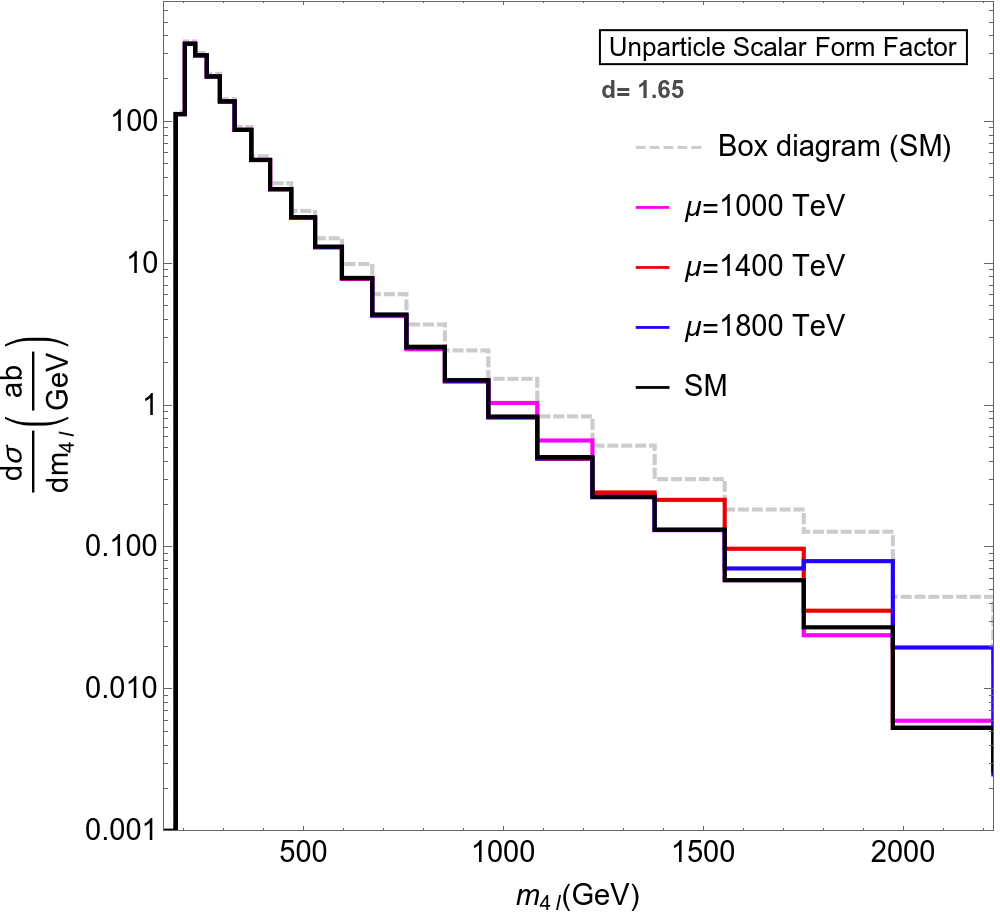}
\caption{Differential cross section as a function of the final $4l$ state invariant mass for the different form factor models of section \ref{sec:other}. The top left figure shows the scalar resonance form factor for the different values of the scalar mixing coupling $\mu$. We assume a scalar width of $10\%$ of the scalar mass. The top right and bottom left figures display the unparticle scalar form factor for a $10\%$ and a $20\%$ widths respectively. The values of the anomalous dimension $d<1.6$ correspond to an enhancement of the coupling below at low energies. In the bottom right figure we show the unparticle case for $d=1.65$, when the form factor induces a suppression of the Higgs coupling. We note that even for a suppressed Higgs coupling, it is still possible to get an enhanced signal with respect to the SM.}
\label{fig:diffcsect_scalar}
\end{center}
\end{figure}

Specifically for  the models presented in Section \ref{sec:other},  there can be an  enhancement of  the SM contribution by enhancing or suppressing the Higgs coupling. For instance, the scalar form factor induced by mixing with a heavy scalar will enhance the coupling for momenta below its pole. On the other hand, in the unparticle case  there will be different possible results  depending on the value of the dimension $d$  of the unparticle operator. For lower values of $d$, we get a structure similar to the simple scalar resonance but with a lower enhancement for increasing $d$. But for larger $d$, the form factor suppresses the coupling in the time-like region. For this case, we can explore the interference structure of the $pp\rightarrow 4l$ to get an enhanced signal. We show the effects of the various form factors in Figure \ref{fig:diffcsect_scalar}. 

Due to the requirement of the Higgs off shellness, the shape modification occurs at the tails of the distributions. Comparing the scalar case  with the unparticle, we observe that the effect of increasing $d$ is similar to a strongly coupled or broader scalar resonance. Additionally, the suppression of the coupling caused by higher values of $d$ for the unparticle case slightly enhances the signal. This degeneracy between the observables of the different form factors should be resolved at higher energies when it is possible to reach the entire new physics sector. The form factors are helpful tools to parametrize the effects at an intermediate energy range when the momentum dependence is accessible through a sufficiently off shell probe.

\section{Conclusions}
\label{sec:conc}

The LHC will be testing the Higgs sector of the SM with increasing precision, particularly at the HL-LHC.
Under these conditions it is necessary to make full use of the increased luminosity even as the LHC remains basically at the same center of mass energy.
In fact, the energy frontier will remain there for years to come. 
Then, it is necessary to consider that the new physics beyond the SM, if not too far above the TeV scale, might appear as modifications of the Higgs couplings.
In this paper, we considered the momentum dependent effects in these couplings. We have taken a model dependent approach, unlike the one taken by effective field theories such as SMEFT~\cite{Grzadkowski:2010es}
and HEFT~\cite{Brivio_2014}. What we loose in generality by considering specific, albeit well motivated, models we gain in our ability to make simple predictions more amenable to comparison with experimental data. 
Unlike the expansions in large numbers of local operators with undetermined coefficients, our approach leads us to  consider models that provide the full non-local physics resulting in the momentum dependence of
the Higgs couplings. 

In order for momentum dependent effects in the Higgs couplings to appear, it is necessary that either the Higgs or another particle participating in the vertex,  be significantly off-shell.
We studied three different models. In Section~\ref{sec:mchm}, we showed that  the minimal CHM results in form factors due to the presence of TeV resonances coupled to the Higgs. The fermion resonances result in a monopole form factor
for the top Yukawa, which can give an enhancement in the time-like region. On the other hand, the vector resonances appear as a dipole, which result in a more subdued effect in the Higgs coupling to the SM massive gauge bosons. The off-shellness of interest is then in the top quarks coupled to the Higgs. We showed in Section~\ref{sec:coll} that $pp\to t\bar t h$ can be sensitive to these effects, as we see in Figure~\ref{fig:cs_tth}. We see that, although we recover the on-shell suppression typical of CHMs, once the momentum dependent effects are taken into account they may result in important enhancements. The same can be said for the $pp \to Vh$ channel, with $V=Z,W^\pm$, although the effects here are much smaller due to the additional suppression mentioned above.

In Section~\ref{sec:other} we studied the form factors generated by the mixing of the Higgs with a heavy scalar, as well as with an unparticle scalar. In these cases, channels with off-shellnes in the Higgs line are crucial in order to be sensitive to the momentum dependent effects. Thus, in Section~\ref{sec:coll} we study the kinematic distributions of $pp \to h^{(*)} \to Z Z \to 4\ell$ as a way to observe these physics, as it can be seen in Figure~\ref{fig:diffcsect_scalar}.
Although these channels have been already considered  elsewhere in connection with momentum dependent effects in Higgs couplings, the treatment presented here may be a first step to provide a map from signals in them to their origins in specific
extensions of the SM.

The increased precision in the measurements of these observables will make it impossible to ignore the momentum dependence of Higgs couplings when considering physics beyond the SM not too far above the TeV scale.in more detailed studies of the We expect to perform more detailed studies of these off-shell signals, at the LHC in run 3, the HL-LHC and future facilities.

\acknowledgments
The authors thank Dorival Gonçalves for helpful discussions. 
They also acknowledge 
the financial support of FAPESP, CNPq and CAPES.

\appendix

\section{MCHM$_\mathbf{5}$ Definitions}
\label{sec:MCHM5gen}
The generators of $SO(5)$ in the fundamental representation can be written as
\begin{eqnarray}
T_{ij}^{a_{L,R}} &=& -\frac{i}{2} \left\{ \frac{1}{2} \epsilon^{abc} \left(\delta_i^b\delta^c_j -\delta^b_j\delta^c_i\right) \pm \left( \delta^a_i\delta^4_j-\delta^a_j\delta^4_i\right)\right\}
  \nonumber\\
& &  \label{eq:generators}\\
  T^{\hat{a}}_{ij} &=& -\frac{i}{\sqrt{2}}\left(\delta^{\hat{a}}_i\delta^5_j-\delta^{\hat{a}}_j\delta^5_i\right)\nonumber~,
\end{eqnarray}
where $i,j=1,\dots,5$, $T^{\hat{a}} $ ($\hat{a}=1,\cdots,4$) are the generators of $SO(5)/SO(4)$, and $T_{ij}^{a_{L,R}} $ with $a_{L,R}=1,2,3$ are the generators of $SO(4)\sim SU(2)_L\times SU(2)_R$.
The spinorial representation of $SO(5)$ can be defined by the matrices:
\begin{equation}
  \Gamma^{\hat{a}} = \begin{pmatrix}
    0 & \sigma^{\hat{a}}\\
    \sigma^{\hat{a}\dagger} & 0
  \end{pmatrix}~, \qquad
  \Gamma^5= \begin{pmatrix}
    \mathbf{1} & 0 \\
    0 & -\mathbf{1}
  \end{pmatrix}~,\qquad
  \sigma^{\hat{a}} = \left\{ \vec{\sigma}, -i\mathbf{1}\right\}~.
\end{equation}

In the unitary gauge, $h^{\hat{a}}=(0,0,0,h)$, the Goldstone matrix is given by
\begin{equation}
U[h] = 
	\begin{pmatrix}
		1& 0& 0& 0& 0 \\
		0& 1& 0& 0& 0 \\
		0& 0& 1& 0& 0 \\
		0& 0& 0& \cos(h/f)& \sin(h/f) \\
		0& 0& 0& \sin(h/f)& \cos(h/f)
	\end{pmatrix}.
      \end{equation}

      Since the elementary quarks are in incomplete $SO(5)$ representations transforming as $SO(4)$ ones, it is convenient to dress the source fields with the NGB matrix. This way we guarantee a fully SO(5) preserving lagrangian, as prescribed by the CCWZ construction.
\begin{align}
Q_l^{\mathbf{[5]}} \equiv U[\Pi(x)] q_L^{\mathbf{5}}, \qquad T_r^{\mathbf{[5]}} \equiv U[\Pi(x)] t_r^{\mathbf{5}}.
\end{align}

Explicit computation in the unitary gauge yields
\begin{equation}
Q_l^{\mathbf{[5]}} =q_l \begin{pmatrix}
0 \\
0 \\
0 \\
\cos(h/f)\\
\sin(h/f)
\end{pmatrix}, \qquad 
T_r^{\mathbf{[5]}} =t_r \begin{pmatrix}
0 \\
0 \\
0 \\
\sin(h/f)\\
\cos(h/f)
\end{pmatrix}.
\end{equation}

\section{Basis Equivalence in Vector Meson Dominance }
\label{sec:basis}
There are different ways of formulating vector meson dominance (VMD). We can introduce mixing interactions between the elementary and composite sectors by either a kinetic  mixing (VMD1) or a mass mixing (VMD2) term.
\begin{equation}
\mathcal{L}_{\gamma\rho}^{\text{VMD1}}=-\frac{e}{2g_\rho}F_{\mu\nu}\rho^{\mu\nu},\qquad\quad \mathcal{L}_{\gamma\rho}^{\text{VMD2}}=-\frac{e m_\rho^2}{g_\rho}\rho_\mu A^\mu.
\end{equation}
The mass mixing in VMD2 can be obtained from HLS. We can use the CCWZ lagrangian with a dynamical $\rho$ resonance to reproduce VMD2~\cite{Bellazzini:2015cgj2}. So, we conclude that HLS, CCWZ and VMD2 are identical descriptions as they produce the same low energy effective theory. In Section \ref{sec:mchm}, we used a kinetic mixing interaction to obtain the Higgs-Gauge boson form factor. This Appendix is dedicated to showing the equivalence of VMD1 with the other approaches more commonly used in the literature. We stress that VMD1 provides a clearer picture of the form factor structure as it maintains explicit gauge-invariant operators in the lagrangian. En route to deriving the equivalence between the two approaches, we show that an interpretation for choosing $a=2$ in HLS can be related to maintaining the gauge invariance of the theory.

In the context of the pion form EM form factor, the VMD2 lagrangian is
\begin{align}
\label{eq:VMD2_lag}\mathcal{L}_{\text{VMD2}}=&-\frac{1}{4}F_{\mu\nu}F^{\mu\nu}-\frac{1}{4}B_{\mu\nu}B^{\mu\nu}+\frac{1}{2}m_\rho^2 \rho_\mu^a \rho^{a,\mu}\\
\nonumber &-\frac{e m_\rho^2}{g} \rho_\mu^3 B^\mu + \frac{1}{2}\left(\frac{e}{g_\rho}\right)^2 m_\rho^2 B_\mu B^\mu -g_{\rho\pi\pi} \rho_\mu J^\mu.
\end{align}
We have a mass mixing term between the composite and elementary gauge bosons. There is no direct coupling between the photon and the hadronic pion current, ensuring the dominance of the $\rho$ meson. Also, we have a mass term for the photon. Even though there is an explicit mass term, the dressed propagator provides the correct pole behavior for $q^2\rightarrow 0$ after summing over all the three-level $\rho$ contributions.
\begin{align}
iD(q^2)&= \includegraphics[scale=0.13]{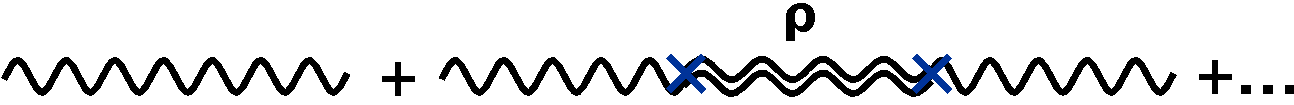}\\
&=\frac{-i}{q^2 -\frac{e^2 m_\rho^2}{g_\rho^2}}+\frac{-i}{q^2 -\frac{e^2 m_\rho^2}{g_\rho^2}}\left(\frac{-ie m_\rho^2}{g_\rho}\right)\frac{i}{q^2-m_\rho^2}\left(\frac{-ie m_\rho^2}{g_\rho}\right)\frac{-i}{q^2 -\frac{e^2 m_\rho^2}{g_\rho^2}}+\dots\\
&=\frac{-i}{q^2-\frac{e^2m_\rho^2}{g_\rho^2}+\frac{e^2 m_\rho^2}{g_\rho^2}\left(\frac{1}{1-q^2/m^2}\right)}\xrightarrow{\mathbf{q^2\rightarrow 0}}\frac{-i}{q^2(1+e^2/g_\rho^2)}.
\end{align}
After rescaling the coupling $e^2\rightarrow e^2(1-e^2/g_\rho^2)$, we arrive at the correct photon propagator with a zero pole mass. So, in VMD2, we introduce the elementary sector without explicit gauge invariance in the lagrangian terms, but the physical quantities are gauge invariant. This procedure makes the obtention of form factors more cumbersome in VMD2. For this reason, it is clearer to include the $\rho$ meson mixings in VMD1, which ensures explicit gauge invariance in the lagrangian directly. 

We can go from VMD2 to VMD1 representations by the following field redefinition:
\begin{align*}
\begin{cases}
\rho_\mu =\rho'_\mu+\frac{e}{g_\rho}A'_\mu,\\
A_\mu = A'_\mu \sqrt{1-e'^2/g_\rho^2},\\
e = e' \sqrt{1-e'^2/g_\rho^2}.
\end{cases}
\end{align*}
The resulting lagrangian for VMD1 is
\begin{align}
\mathcal{L}_{\text{VMD1}}=&-\frac{1}{4}F'_{\mu\nu}F'^{\mu\nu}-\frac{1}{4}\rho'_{\mu\nu}\rho'^{\mu\nu} +\frac{1}{2}m_\rho^2\rho'_\mu \rho'^\mu \\
\nonumber & -g_{\rho\pi\pi} \rho_\mu J^\mu - e A'_\mu J^\mu -\frac{e}{2g_\rho}F'_{\mu\nu}\rho'^{\mu\nu}.
\end{align}
So, we can see that on this basis, the photon mixes kinetically with the $\rho$ meson. Additionally, there is a direct coupling between the photon and the pion hadronic current, and the undressed photon is massless with explicit gauge-invariant terms. Kinetic mixing also provides a clearer form factor normalisation since the quadratic term is diagonal for an on-shell photon. This structure means that the photon decouples from $\rho$ at zero momenta, $q^2\rightarrow 0$.
\begin{equation}
F'_{\mu\nu}\rho'^{\mu\nu}=2 q^2 A'_\mu \rho'^\mu \xrightarrow{\text{on-shell}} 0
\end{equation}

In addition to VMD1 and VMD2, we can define the decoupled basis in which the quadratic terms are diagonalized. Redefining the fields from \eqref{eq:VMD2_lag} as
\begin{align}
\begin{cases}
B_\mu=\frac{1}{\sqrt{g_\rho^2+e^2}}(g A_\mu +e \rho_\mu^3)\\
V_\mu=\frac{1}{\sqrt{g_\rho^2+e^2}}(g \rho_\mu^3-e A_\mu),
\end{cases}
\end{align}
leads to
\begin{align}
\mathcal{L}_{\text{diag}}=&-\frac{1}{4}\left[B_{\mu\nu}B^{\mu\nu}+V_{\mu\nu}V^{\mu\nu}\right]+\frac{1}{2}m_\rho^2 V_\mu V^\mu -\left(g_{\rho\pi\pi} V_\mu + e B_\mu\right)J^\mu.
\end{align}

\begin{equation}
\label{eq:ffVMD1}F_\pi^{\text{VMD1}}(q^2)=1-\frac{g_{\rho\pi\pi}}{g_\rho}\frac{q^2}{q^2-m_\rho^2}.
\end{equation}
The boundary conditions for the form factor ensures the universality condition $g_{\rho\pi\pi}=g_\rho$.
\begin{align}
\begin{cases}
F_\pi(0)=1\\
F_\pi(\infty)=0 \quad\Rightarrow\quad g_{\rho\pi\pi}=g_\rho
\end{cases}
\end{align}
\begin{figure}
\centering
\includegraphics[scale=0.3]{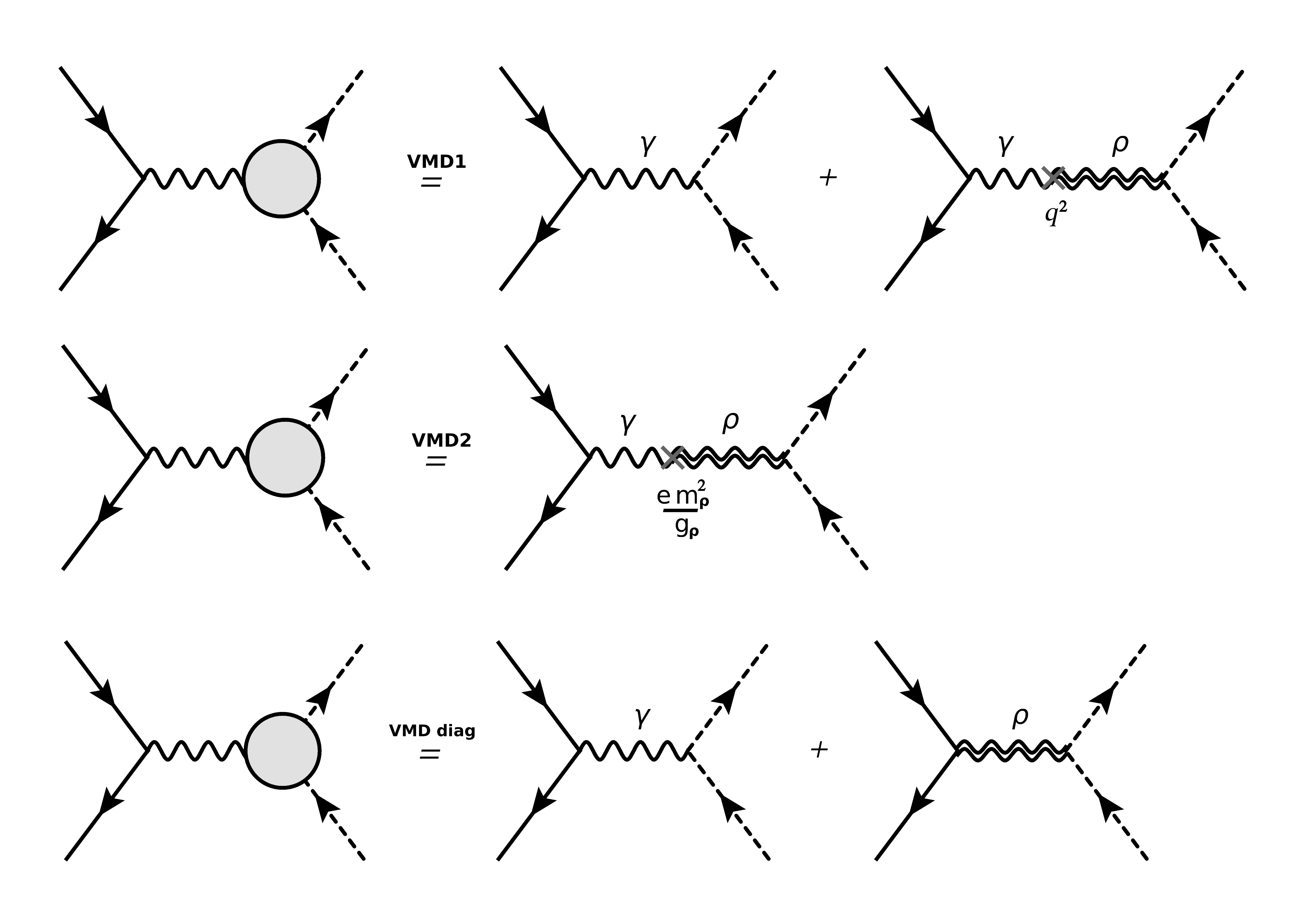}
\caption{Pion form factor in the different Vector Meson Dominance basis.}
\label{fig:VMDcomb}
\end{figure}

There is no vector-meson mixing on the physical basis and the photon couples directly to the pion current. So, what happened to the $\rho$-dominance of the EM current? As argued in~\cite{Schechter:1986vs}, the dominance of the $\rho$ meson is entirely basis dependent, but the pion form factor is the same in all formulations. The inclusion of gauge interactions between elementary and composite sectors is independent of the basis on which the composite sector is introduced. Accordingly, throughout Section~\ref{sec:mchm}, we have used this result from HLS - the theory is independent of the basis on which the dynamical gauge boson is introduced.

Before proceeding tho the MCHM$_\mathbf{5}$ example, it is useful to show explicitly that the pion form factor is the same in all VMD basis discussed so far. The form factor appears to correct the elementary electron-positron scattering to pions.
\begin{equation}
\mathcal{M}_{\gamma\rightarrow\pi^+\pi^-}^\mu=-e(p^+-p^-)^\mu F_\pi(q^2).
\end{equation}
The form factors are obtained for each basis as shown in Figure \ref{fig:VMDcomb}. In VMD1, the form factor is the sum of the direct photon-pion terms and the $q^2$ mixing with the $\rho$.

In VMD2, there is no direct pion coupling to the photon and the elementary electron only couples to the photon. The form factor is then
\begin{equation}
\label{eq:ffVMD2}F_\pi^{\text{VMD2}}(q^2)=\frac{-m_\rho^2}{q^2-m_\rho^2}\frac{g_{\rho\pi\pi}}{g_\rho},
\end{equation}
and the universality condition is satisfied by the on-shell normalization $F_\pi(0)=1$. The asymptotic condition $F_\pi(\infty)=0$ is satisfied automatically. 

In the diagonal basis, the electrons are partially composite and couples to the $\rho$ meson. The form factor is the sum of both the photon and $\rho$ s-channel contributions.
\begin{equation}
\label{eq:ffVMDdiag}F_\pi^{\text{diag}}(q^2)=1-\frac{g_{\rho\pi\pi}}{g_\rho}\frac{q^2}{q^2-m_\rho^2}
\end{equation}
Thus, we conclude that after imposing the appropriate boundary conditions on the expressions, all form factors are identical\footnote{Note that \eqref{eq:ffVMD1} and \eqref{eq:ffVMDdiag} are equal to \eqref{eq:ffVMD2} after imposing the universality condition.}. 

\subsection*{Composite Higgs Gauge Form Factor}
Shifting to the form factors in composite Higgs models, we can derive the composite sector lagrangian as in the HLS approach. This method lead to a mass mixing as in VMD2. The CCWZ construction permits two invariants,
\begin{align}
&\mathcal{L}_1=\frac{f^2}{4}\Tr\left(d_\mu d^\mu\right),\\
&\mathcal{L}_2=\frac{f^2}{4}\Tr\left(E_\mu E^\mu\right)=\frac{f^2}{4}\Tr(g_\rho \rho_\mu-e_\mu)^2.
\end{align}
where the CCWZ symbols $d^\mu$ and $e^\mu$ are defined from $U\partial_\mu U=d_\mu^{\hat{a}}T^{\hat{a}}+e^a_\mu T^a$. The composite sector lagrangian is
\begin{equation}
\mathcal{L}_{CS}=\mathcal{L}_1+a\mathcal{L}_2 -\frac{1}{4}\rho^a_{\mu\nu} \rho^{a,\mu\nu},
\end{equation}
where $a$ is the HLS phenomenological constant and assuming that strong dynamics generates a kinetic term for the $\rho$. Expanding the CCWZ symbols, we arrive at
\begin{align}
\label{eq:LCS}\mathcal{L}_{CS}=-\frac{1}{4}\rho^a_{\mu\nu} \rho^{a,\mu\nu}+\frac{f^2 g_\rho^2}{2}\rho_\mu^a \rho^{a,\mu}+ \frac{1}{2}(\partial_\mu h)^2 + g_\rho \rho_\mu^a (h \overleftrightarrow{\partial_\mu} h )^a +\frac{1}{2}g_\rho^2 \rho_\mu^a \rho^{a,\mu} |h|^2,
\end{align}
where $h$ is the Higgs field.

We want to include the local electroweak symmetry as an external gauging of $SU(2)_L\times U(1)_Y$. First, we add the kinetic terms for the EW bosons,
\begin{equation}
-\frac{1}{4} W^{a_L}_{\mu\nu} W^{a_L \mu\nu} -\frac{1}{4} B_{\mu\nu} B^{\mu\nu},
\end{equation}
where the SM field are embedded in $SO(4)\simeq SU(2)_L\times SU(2)_R \supset SU(2)_L\times U(1)_Y$ as $A_\mu^{a_L}=W^a_\mu$ and $A^{3R}_\mu=B_\mu$. The mass terms and elementary composite mixing interactions arise from 
\begin{align}
\frac{f^2}{4}\Tr(g_\rho \rho_\mu-e_\mu)^2 \,\supset\, & \frac{g^2 f^2}{2} W^{a_L}_{\mu} W^{a_L \mu}+\frac{g'^2 f^2}{2}B_\mu B^\mu \\
\nonumber & -f^2 g_\rho g \rho_\mu^{a_L} W^{a_L, \mu} - f^2 g_\rho g' \rho^{3R}_\mu B^\mu\\
\nonumber & - g_\rho g \rho_\mu^{a_L} W^{a_L, \mu}|h|^2 - g_\rho g' \rho^{3R}_\mu B^\mu |h|^2
\end{align}
Thus, the lagrangian of the elementary sector is
\begin{equation}
\label{eq:LES}\mathcal{L}_{ES}=-\frac{1}{4} W^{a_L}_{\mu\nu} W^{a_L \mu\nu} -\frac{1}{4} B_{\mu\nu} B^{\mu\nu} + \frac{g^2 f^2}{2} W^{a_L}_{\mu} W^{a_L \mu}+\frac{g'^2 f^2}{2}B_\mu B^\mu 
\end{equation}
and the mixing interactions are
\begin{equation}
\label{eq:Lmix}\mathcal{L}_{mix}=-f^2 g_\rho g \rho_\mu^{a_L} W^{a_L, \mu} - f^2 g_\rho g' \rho^{3R}_\mu B^\mu - g_\rho g \rho_\mu^{a_L} W^{a_L, \mu}|h|^2 - g_\rho g' \rho^{3R}_\mu B^\mu |h|^2
\end{equation}

As in the pion VMD2 formulation, there are several features to note on \eqref{eq:LCS}, \eqref{eq:LES} and \eqref{eq:Lmix}. First, the undressed EW fields are massive\footnote{This is before introducing the Higgs mechanism for EWSB.}, but the sum of the one particle-irreducible contributions to the $\rho$-mixing terms guarantees the correct $q\rightarrow 0$ pole behavior.
\begin{align}
D(q^2)&=\left(q^2 -\frac{g^2 m_\rho^2}{g_\rho^2}-\frac{e^2 m_\rho^4}{g_\rho^2(q^2-m_\rho^2)}\right)^{-1}\\
&=\frac{1}{q^2(1+\frac{g^2}{g_\rho^2})} \qquad \text{as $q\rightarrow 0$}.
\end{align}

\begin{figure}
\centering
\includegraphics[scale=0.04]{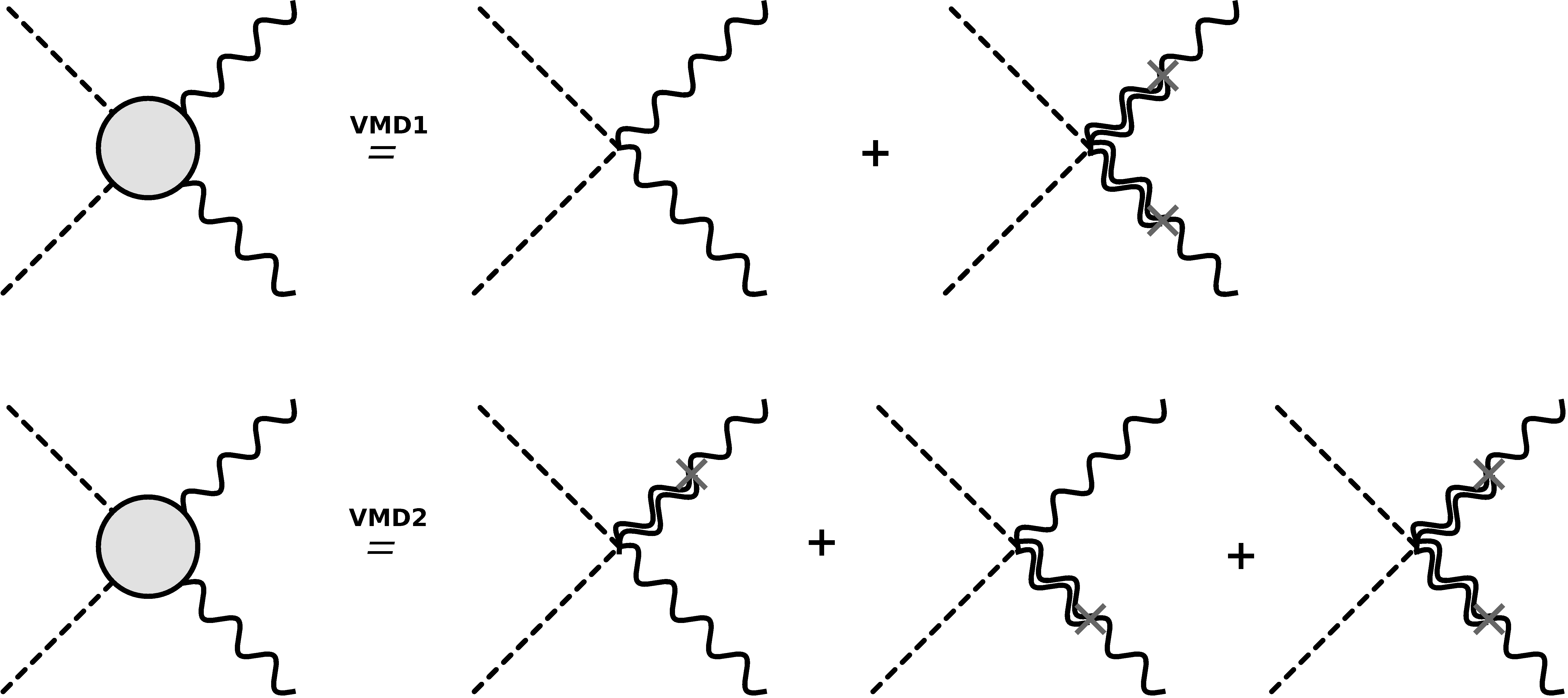}
\caption{Higgs form factors obtained in the different VMD basis for the MCHM$_\mathbf{5}$.}
\label{fig:higgsVMD}
\end{figure}

Note that the additional momentum dependence of $D(q^2)$ constitutes a form factor for the EW gauge bosons two-point functions. Additionally, the Higgs form factor with the EW bosons is obtained without a direct $A^2$ coupling to the Higgs as in figure \ref{fig:higgsVMD}. The form factor is
\begin{equation}
\label{eq:monopole_dipole_ff}\frac{f(p_1,p_2)}{g^2}=-\frac{m_\rho^4}{(p_1^2-m_\rho^2)(p_2^2-m_\rho^2)}-\frac{m_\rho^2}{p_1^2-m_\rho^2}-\frac{m_\rho^2}{p_2^2-m_\rho^2}.
\end{equation} 
The form factor has a dipole contribution due to the diagram with two $\rho$s and the monopole terms with one $\rho$. The monopole interactions appear because the EW bosons and the $\rho$ interactions with the Higgs are not diagonal on this basis. The term $\rho A h^2$ term does not cancel with $\mathcal{L}_1$ when we choose $a=2$ in HLS as there is no such coupling in $\mathcal{L}_1$. These are features of the mass mixing basis in which we do not maintain explicit gauge invariance in the lagrangian operators.

Our goal now is to change the basis from a mass to a kinetic mixing to reproduce the lagrangian used in Section~\ref{sec:mchm}. Analogously to the process in VMD, we redefine the fields as 
\begin{align}
\begin{cases}
\rho'^{a_L}_\mu=\rho^{a_L}_\mu-\frac{g}{g_\rho}W^{a_L}_\mu \qquad, \rho'^{3_R}_\mu= \rho^{3R}_\mu-\frac{g'}{g_\rho}B_\mu\\
W^{a_L}_\mu = \sqrt{1-\frac{g^2}{g_\rho^2}}W_\mu^{a_L} \qquad, B'_\mu = \sqrt{1+\frac{g'^2}{g_\rho^2}} B_\mu \\
\tilde{g}=-g\sqrt{1+\frac{g^2}{g_\rho^2}} \qquad \tilde{g'}=-g' \sqrt{1+\frac{g'^2}{g_\rho^2}}
\end{cases}
\end{align}    
Finally, the complete lagrangian of the MCHM$_\mathbf{5}$ gauge bosons in the VMD1 basis is
\begin{align}
\mathcal{L}_{CS}^{V}=&-\frac{1}{4}\rho_{\mu\nu}^a \rho^{a,\mu\nu}+\frac{m_\rho^2}{2}\rho^a_{\mu}\rho^{a,\mu}+g_\rho \rho_\mu^a J^{a,\mu}+\frac{g_\rho^2}{2}\rho^a_{\mu}\rho^{a,\mu} h^2,\\
\mathcal{L}_{ES}^{V}=&-\frac{1}{4}W_{\mu\nu}^{a_L} W^{a_L,\mu\nu}+g W_\mu^{a_L} J^{a_L,\mu}+\frac{g^2}{2}W^{a_L}_{\mu}W^{a_L,\mu} h^2\\
&\nonumber -\frac{1}{4}B_{\mu\nu} B^{\mu\nu}+g' B_\mu J^{3_R,\mu}+\frac{g'^2}{2}B_{\mu}B^{\mu} h^2,\\
\label{eq:kinmix3}\mathcal{L}_{int}^{V}=&\frac{1}{2}\frac{g}{g_\rho} W_{\mu\nu}^{a_L} \rho^{a_L,\mu\nu} + \frac{1}{2}\frac{g}{g_\rho} B_{\mu\nu} \rho^{3_R, \mu\nu},
\end{align} 
which is the lagrangian \eqref{eq:kinmix} used in Section~\ref{sec:mchm}. We see that with kinetic mixing, gauge invariance is explicitly maintained with the cancellation of the mass terms for the EW bosons. The vertices that introduced the monopoles in the form factor also disappear, leaving only diagonal gauge couplings as expected by gauge invariance. Note that the matching between the two different basis depends on choosing $a=2$. Without $a=2$, the matching of the basis would not produce a gauge invariant kinetic mixing formulation. In this sense, the phenomenological choice $a=2$ of HLS is justified by requiring a gauge invariant model for the inclusion of the $\rho$. Finally, the Higgs form factor is obtained as in Figure~\ref{fig:higgsVMD}. The expression is 
\begin{equation}
\label{eq:ff_dipole}\frac{f(p_1,p_2)}{g^2}=1-\frac{p_1^2 p_2^2}{(p_1^2-m_\rho^2)(p_2^2-m_\rho^2)}.
\end{equation} 
After inspection, we observe that both expression \eqref{eq:ff_dipole} and \eqref{eq:monopole_dipole_ff} are identical as expected by the basis independence in which the dynamical resonances are introduced.

\bibliography{ffacrefs}
\bibliographystyle{unsrt}

\end{document}